\documentclass[twocolumn,aps,prl,nofootinbib,nolongbibliography]{revtex4-2}

\usepackage[]{graphicx}

\usepackage{amsfonts,amsmath,amssymb}
\usepackage{dcolumn}
\usepackage{bm}
\usepackage{mathrsfs}
\usepackage{bm}
\usepackage[normalem]{ulem}
\usepackage[dvipsnames]{xcolor}

\usepackage[setpagesize=false,
 bookmarksnumbered=true,%
 bookmarksopen=true,%
 colorlinks=true,%
 linkcolor=blue,
 citecolor=blue,
 urlcolor=blue
 ]{hyperref}

\newcommand{\Bra}[1]{\left( #1 \right)}				
\newcommand{\BRa}[1]{\left[ #1 \right]}				
\newcommand{\ket}[1]{| #1 \rangle}			
\newcommand{\bra}[1]{\langle #1 |}			
\newcommand{\bracket}[2]{\langle #1 | #2  \rangle} 	
\newcommand{\ketbra}[2]{| #1 \rangle \! \langle #2 |}
\newcommand{\pdo}[2]{\frac{\partial #1}{\partial #2}}		
\newcommand{\1}{\mbox{1}\hspace{-0.25em}\mbox{l}}	
\newcommand{\eps}{\epsilon}

\newcommand{\tr}[0]{\mathrm{tr}}

\newcommand{\E}[0]{\mathcal{E}}

\newcommand{\eff}[0]{\mathrm{eff}}

\newcommand{\uo}[0]{u^\Gamma}
\newcommand{\Eo}[0]{\mathcal{E}^\Gamma}
\newcommand{\psio}[0]{\Psi^\Gamma}
\newcommand{\psiLo}[0]{\Psi^\Gamma_L}
\newcommand{\A}[0]{\mathcal{A}}
\newcommand{\nsep}[0]{n_\mathrm{sep}}

\def\<{\langle}
\def\>{\rangle}

\usepackage{cancel}

\begin{document}

\title{Dynamical tunneling across the separatrix}
 \author{Yasutaka Hanada}
 \email{hanada@cas.showa-u.ac.jp}
 \affiliation{Department of Information Science, Showa University, Yamanashi 403-0005, Japan}
 \affiliation{Department of Physics, Tokyo Metropolitan University, Tokyo 192-0397, Japan}
 \author{Kensuke S. Ikeda}
 \affiliation{Department of Physics, Ritsumeikan University, Kusatsu, Shiga, 525-0577, Japan}
 \author{Akira Shudo}
 \affiliation{Department of Physics, Tokyo Metropolitan University, Tokyo 192-0397, Japan}

\date{\today}
\begin{abstract}

The strong enhancement of tunneling couplings typically observed in tunneling splittings in the quantum map is investigated. 
We show that the transition from instanton to noninstanton tunneling, which is known to occur in tunneling splittings in the space of the inverse Planck constant, takes place in a parameter space as well. 
By applying the absorbing perturbation technique, we find that the enhancement invoked as a result of local avoided crossings and that originating from globally spread interactions over many states should be distinguished and that the latter is responsible for the strong and persistent enhancement.  
We also provide evidence showing that the coupling across the separatrix in phase-space is crucial in explaining the behavior of tunneling splittings by performing the wave-function-based observation. 
In the light of these findings, we examine the validity of the resonance-assisted tunneling theory.

\end{abstract}
\maketitle
\section{Introduction}

Tunneling splitting is a typical manifestation of the tunneling effect in quantum mechanics and it is observed, for example,  in a system with a symmetric double-well potential. 
The states with identical energy but different parities are quasi degenerate and exhibit exponentially small energy splittings, which occur as a result of the tunneling coupling between two states supported by energetically separated potential wells. 

Although the tunneling splitting is a purely quantum effect, it is possible to evaluate the splitting width in terms of the classical orbit if one is allowed to use the complex plane.  
The so-called instanton is a complex classical orbit connecting two valleys, and its classical action is known to provide the splitting width \cite{simon1983,simon1984,coleman1988}. 
There are rigorous mathematical results that claim that the tunneling splitting $|E_1-E_0|$ for the ground-state doublet behaves as 
\begin{eqnarray}
\lim_{\hbar \to 0} \Bigl( - \hbar \ln |E_1-E_0| \Bigr) = \rho (a,b), 
\end{eqnarray}
where $\rho(a,b)$ is the Agmon distance between the bottoms $a$ and $b$ of the potential well and the Agmon distance corresponds to the classical action of the instanton orbit. Note that the above estimate holds not only for one-dimensional but also for multi-dimensional systems. 

In one-dimensional systems, for the energy above the potential barrier, the two wells are connected via classical orbits in the real plane so that tunneling doublets disappear. 
On the other hand, in multi-dimensional systems, additional conserved quantities other than energy may exist in the system, either locally or globally. In that case, 
classical orbits may be confined in one of the potential wells, even though the energy of the orbits is 
above the potential barrier. This means that the transition between the potential wells is energetically allowed but dynamically forbidden. 
The quantum state associated with classical orbits dynamically confined in one of the wells is sometimes called local modes, and tunneling splittings appear between (symmetrically) formed local modes as well as tunneling splittings created between the states below the potential barrier. Such local modes are coupled via the tunneling effect since their supports are classically separated, and this type of tunneling is called dynamical tunneling~\cite{davis1981,keshavamurthy2011}. 

Evaluation of the width of tunneling splittings associated with dynamical tunneling would be beyond the standard approach using the instanton path. The difficulty of this problem lies in the fact that the system is nonintegrable in general and chaos appears in the corresponding classical dynamics. 

As reported in the literature~\cite{roncaglia1994,bonci1998,brodier2001,brodier2002,mouchet2007,lock2010,keshavamurthy2011,hanada2015,hanada2019} and indeed demonstrated in this paper, tunneling splittings do not follow a simple exponential law but rather change their slope or even show plateau-like structures in the splitting width versus the inverse Planck constant plot.
This implies that the magnitude of tunneling couplings induced by dynamical tunneling can be much larger than those predicted from the instanton or the Agmon distance argument. 
The underlying mechanism of dynamical tunneling is thus expected to be completely different from that derived from the existing theory. 

Although much effort has been devoted to clarifying the signature of tunneling splittings, focusing mainly on the influence of nonintegrability~\cite{bohigas1993a,bohigas1993,brodier2002,keshavamurthy2011,shudo2014,hanada2015,hanada2019,iijima2022}, we have to say that our understanding is still in the dark. 
This is mainly due to the lack of an explicit or closed semiclassical formula that allows us to evaluate tunneling splittings directly. The problem is considered to be almost insurmountable 
because of enormously complex aspects of the underlying classical dynamics. Generic systems, in which dynamical tunneling occurs, 
are neither completely integrable nor fully chaotic, so in contrast to Gutzwiller's trace formula for ideally chaotic systems, semiclassical formulas for eigenvalues and eigenfunctions are not known in any form. 
The best we can do would be to apply the WKB method in the 
time domain, for which an explicit semiclassical formula is available. 
The result reveals that chaos in the complex plane plays a crucial role in the transition between classically forbidden phase-space regions~\cite{shudo1995complex,shudo1998chaotic,shudo2002julia,shudo2009a,shudo2009b}. 
However, the connection between the behavior of tunneling splittings and chaos in the complex plane has yet to be clarified.

Under these circumstances, several possible scenarios have been proposed on a phenomenological level to capture the signature of tunneling in nonintegrable systems. 
The analyses intend to understand how classical chaos 
and other phase-space structures relate to the 
behavior in nonintegrable tunneling. 
The present paper is essentially along the same lines 
or even more phenomenological. 
Here our strategy is to observe closely the nature of the tunneling couplings under suitably chosen bases, which are given by applying the Baker-Campbell-Hausdorff (BCH) formula. 
Through such purely quantum mechanical analyses, we explain the mechanism leading to the strong enhancement of the tunneling probability, which is a typical feature in the nonintegrable systems~\cite{shudo2014,hanada2015,hanada2019,iijima2022}. 

Among many other bases representing quantum states, 
the BCH basis provides a certain privilege because it can sharply 
capture the transition from the instanton to non instanton (INI) tunneling~\cite{shudo2014}. 
Here instanton tunneling refers to a type of tunneling for which the instanton approximation works in evaluating tunneling couplings. In contrast, non-instanton tunneling is that which the instanton approximation cannot describe. 

As reported in Refs.~\cite{hanada2015,hanada2019}, and will be shown in this paper, 
tunneling couplings have unexpectedly broad supports 
in the BCH representation, and the strong enhancement is caused by 
couplings across the separatrix in the phase-space. 
This picture contrasts sharply with the so-called 
resonance-assisted tunneling (RAT) scenario 
because the RAT theory incorporates couplings only inside the separatrix.  
In the RAT theory, the coupling via nonlinear resonances is introduced 
by constructing a local integrable Hamiltonian, thereby tunneling 
couplings in the RAT theory basically originate from the associated couplings 
derived in the integrable system. 
In other words, the coupling can be introduced only
between the regions connected by the instanton,
so the transition across the separatrix is clearly beyond the scope of the RAT theory.

The existence of broad supports in the BCH basis explains why the strong coupling persists even when one varies the Planck constant or a system parameter. At the same time, it provides a reason why, in the RAT calculation, one must retain the couplings associated with classical nonlinear resonances even when the resonance condition no longer holds. 
Alternatively stated, if one includes the RAT coupling only when the resonance condition is satisfied, one cannot reproduce the characteristic signatures in the splitting plots.

The organization of the paper is as follows.
In Sec.~\ref{sec:preparations} we introduce the system used here to study the nature of tunneling splittings. 
We consider a two-dimensional area-preserving map, which is regarded as a model of the Poincar\'e surface of section of the two-dimensional continuous flow Hamiltonian system. 
We should consider the extent to which the results obtained in the area-preserving map can be applied to the continuous flow system for which the question of dynamical tunneling was originally posed.  
We then introduce the integrable approximation to the area-preserving 
map. Here the BCH formula is used to give an integrable approximation. 
In Sec. \ref{sec:spl_vs_eps}  we first study the behavior of tunneling splittings as a function of 
the perturbation strength, and show that 
the so-called instanton-noninstanton transition~\cite{shudo2014,hanada2015}
occurs in the splitting plot. 
We then apply the absorbing perturbation technique 
to elucidate that the strong enhancement of the 
tunneling couplings after the INI transition is
supported by broad interactions over many states. 
In Sec. \ref{sec:wave_function_based}
we perform a wave-function-based analysis for tunneling splittings. 
We show that tunneling splittings can be evaluated by 
referring to the amplitude of the wave function at a specific point, 
the center of two symmetrically located regular regions. 
This is done using the Herring-Wilkinson formula, which allows us to 
evaluate the magnitude of the tunneling splittings in terms of wave functions 
in the two dynamically separated regions. 
The wave-function-based analysis reveals that the coupling beyond the separatrix gives rise to
the enhancement of the splittings. Moreover, the coupling with the region outside the separatrix is already present in the wave functions even in the instanton regime, 
although no anomalous signature is apparently visible in the tunneling splittings. 
In Sec. \ref{sec:spl_vs_hbar} we examine the splitting plot as a function of $1/\hbar$, which has often been studied when testing the RAT scenario \cite{brodier2001,brodier2002,keshavamurthy2005b,eltschka2005,mouchet2006,lock2010,schlagheck2011,deunff2013}. 
We again use the absorbing perturbation method to confirm the robustness of the tunneling splitting enhancement, 
and then test the validity of the RAT calculation by actually applying the proposed recipe. 
Based on the result, we point out that the persistent enhancement is not due to the RAT-type interaction but to the interactions widely spread over many levels. Here the persistent enhancement is referred to as a phenomenon in which the anomalous enhancement of tunneling splittings compared to the integrable limit persists even with the change of $1/\hbar$.
 Section \ref{sec:Conclusion} provides a summary and discussion.

\section{preparations}
\label{sec:preparations}

\subsection{CLASSICAL AND QUANTUM SYSTEMS}

In this paper, we use the kicked-rotator Hamiltonian
\begin{equation}\label{eq:kicked-ham}
 H(p,q,t) = T(p) + \epsilon V(q)\sum_{n \in \mathbb{Z}}\delta(t-n\tau),
\end{equation}
as a model of non-integrable systems,
where $\epsilon$ and $\tau$ are the strength and period of a perturbation, respectively.
The angular frequency of the perturbation is defined as $\Omega = 2\pi/\tau$.

By using a half-$\tau$ (symmetrized) integration,
the classical dynamics from the $n$-th kick to the $(n+1)$-th kick is expressed as the symplectic mapping
\begin{equation}\label{eq:cmap}
f=f_V(\tau/2)\circ f_T(\tau)\circ f_V(\tau/2),
\end{equation}
where $f_V(\tau):(q,p)\mapsto (q, p -\tau\eps V'(q))$ and $f_T(\tau):(q,p)\mapsto(q+\tau T'(p),q)$.
Here the prime stands for the derivative with respect to the argument.
The classical map $f$ is equivalent to the second-order symplectic numerical integrator (scheme)
for the autonomous Hamiltonian $H_1(p,q) = T(p) + \epsilon V(q)$ with a time-step size $\tau$. 

\begin{figure}
  \centering
  \includegraphics[width=0.48\textwidth]{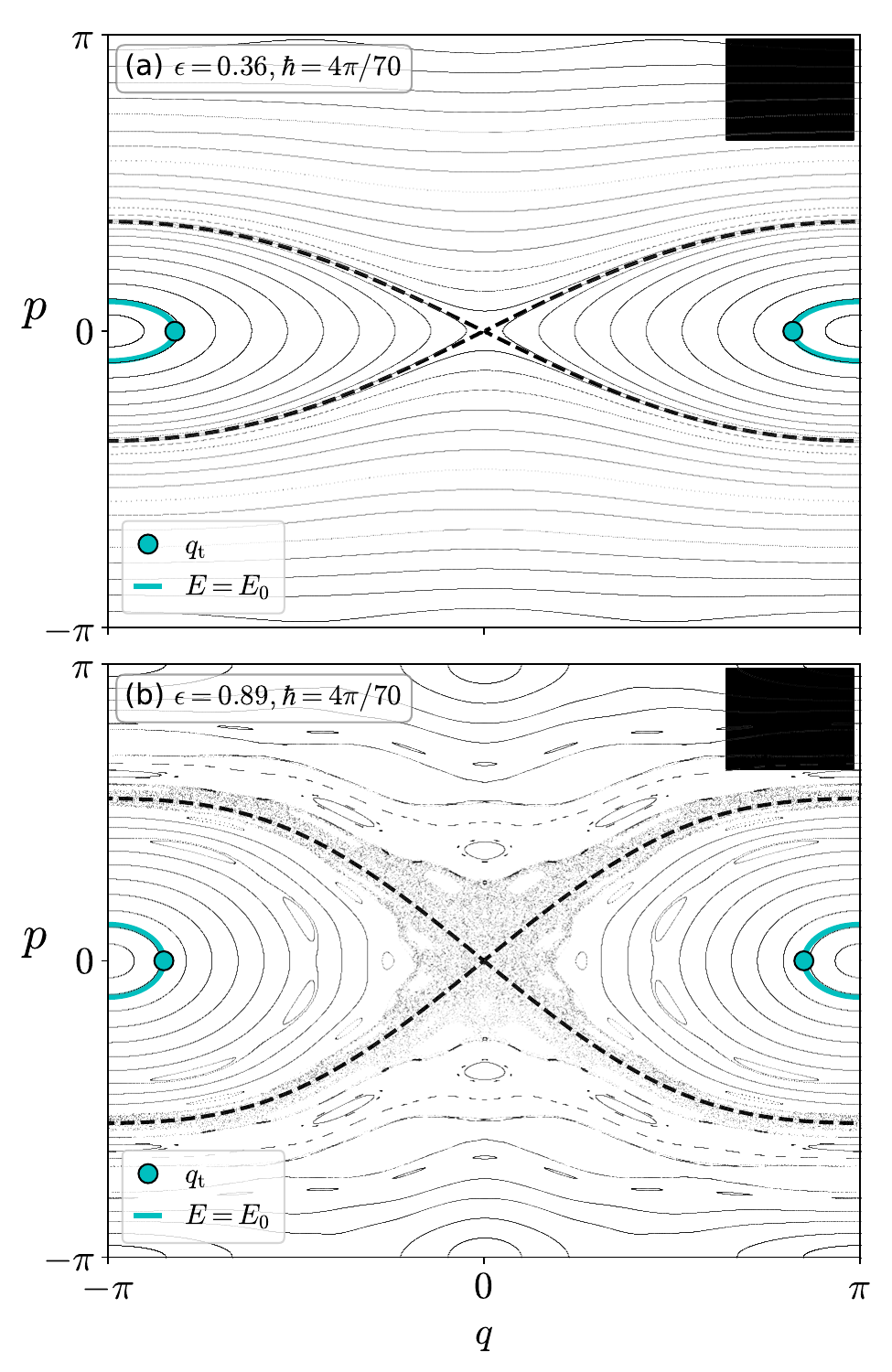}
  \caption{\label{fig:stan_ps}
      Plot of $p$ vs $q$ for (a) $\epsilon=0.36$ and (b) $\epsilon=0.89$. Gray dots represent the points generated by the classical dynamics $f$.
      The cyan solid curve and dots show the energy contours of the classical BCH Hamiltonian $H_\mathrm{cl}^{(M)}$ $(M=7)$ related to $E_0$ and the associated turning points $q_\mathrm{t}$ for $\hbar=4\pi/70$, respectively.
      The black broken curve represents the separatrix $E=\eps$ of the $H_\mathrm{cl}^{(M)}(p,q)$. The black box in (a) and (b) represents the effective Plank cell.}
\end{figure}

In this paper, we set $T(p)=p^2/2$, $V(q)=\cos q$, and $\tau=1$.
The classical map $f$ is called the (symmetrized) Chirikov-Taylor standard map.
Typical phase-space portraits generated by $f$ are shown in Figs.~\ref{fig:stan_ps}(a) and \ref{fig:stan_ps}(b).

Adopting the canonical quantization,
the wave packet dynamics from the $n$-th kick to the ($n+1$)-th kick is expressed as
\begin{equation}\label{eq:qmap}
 \hat{U} = e^{-\frac{i}{\hbar}\frac{\epsilon}{2}V(\hat{q})}
e^{-\frac{i}{\hbar}T(\hat{p})}
e^{-\frac{i}{\hbar}\frac{\epsilon}{2}V(\hat{q})},
\end{equation}
which is referred to as the quantum map~\cite{berry1979,casati1979}.
We focus our attention on quasi-stationary states of the quantum map (\ref{eq:qmap}).
The eigenvalue equation is given as
\begin{equation}\label{eq:qmap_eigen}
  \hat{U}\ket{\Psi_n} = u_n\ket{\Psi_n},\qquad u_n = e^{-\frac{i}{\hbar}\E_n},
\end{equation}
where $\ket{\Psi_n}$ is the quasi eigenstate (Floquet state) and $\E_n$ is the associated quasi-eigenenergy.

Below we assign a quantum number in ascending order of the eigenvalues for $H_1$:
The quantum number $m$ of eigenstates $\ket{\Psi_m}$ is 
determined by the quantum number of an eigenstate $\ket{J_n^{(1)}}$ of $H_1$ attaining  
the maximal overlap $|\bracket{\Psi_m}{J^{(1)}_n}|^2$. 

\subsection{\uppercase{integrable approximations to the quantum map}}

The quantum map (\ref{eq:qmap})
 is expressed as a product of noncommutative operators. Here we introduce an integrable approximation for the operator $\hat{U}$ by applying the BCH formula~\cite{scharf1988,shudo2014,hanada2015}.
The BCH expansion for the quantum map $\hat{U}$ gives an infinite series of the Hamiltonian
expressed as
\begin{equation}\label{eq:bch_exp}
\hat{H}_\mathrm{eff}(\hat{p},\hat{q}) = \sum_{j\in \text{odd integer}}\Bra{-\frac{i}{\hbar}}^{j-1}\hat{h}_j(\hat{p},\hat{q}),
\end{equation}
with $\hat{h}_j$ given by
\begin{equation}
\begin{split}
\hat{h}_1 &= \hat{T} + \epsilon \hat{V},\quad\\
\hat{h}_3 & = \frac{1}{24}\Bra{[\hat{T},[\hat{T},\epsilon \hat{V}]] - [\epsilon\hat{V},[\epsilon\hat{V},\hat{T}]]}, \\\hat{h}_5 & = \cdots,
\end{split}
\end{equation}
where $[\cdot,\cdot]$ is a commutator.
Here we call $\hat{H}_\mathrm{eff}(\hat{p},\hat{q})$ the effective quantum BCH Hamiltonian.
Assuming the canonical quantization rule $\hat{p}\psi(q)= \frac{\hbar}{i}\pdo{}{q} \psi(q)$ and $\hat{q}\psi(q) = q\psi(q)$,
the Hamiltonian (\ref{eq:bch_exp}) can be rewritten as a power series of $\hbar$,
\begin{align}\label{eq:bch_ham_hbar}
\hat{H}_\mathrm{eff}(\hat{p},\hat{q})= H_\mathrm{cl}(\hat{p},\hat{q})+O(\hbar).
\end{align}
The classical Hamiltonian $H_\mathrm{cl}$ can be obtained by taking the limit of $\hbar\to0$ in Eq.~(\ref{eq:bch_ham_hbar}).
Note that the classical Hamiltonian $H_\mathrm{cl}$ is equivalent to a modified Hamiltonian for the classical map $f$
in the context of the geometric numerical integration~\cite{hairer2010,faou2012}.

In some situations, it was proved that the BCH series (\ref{eq:bch_ham_hbar}) converge under appropriate operator norms~\cite{blanes2004},
but we are not sure that the current BCH series does so.
Within our numerical calculations,
it was shown that the accuracy of the series improves up to a certain optimal truncation order and
provides an extremely good approximation to the exact results,
but the series becomes less accurate as the order increases further~\cite{hanada2015}.
The series thus behaves like an asymptotic series.
On the other hand, if we regard the BCH series (\ref{eq:bch_ham_hbar}) as a classical Hamiltonian,
the resulting Hamiltonian flows trace well the iterated orbits generated by the classical map $f$
if the order of truncation is taken to be optimal.
However, it is known that the classical BCH series does not converge in general~\cite{hairer2010}.

Throughout our paper, we use the truncated BCH Hamiltonian $H_\eff^{(M)}$, where $M$ stands for the order of truncation in the series (\ref{eq:bch_exp}).
Now we introduce the eigenvalue $E_n^{(M)}$ and the corresponding eigenstate $\ket{J_n^{(M)}}$ for $\hat{H}^{(M)}_\eff$ as
\begin{equation}
  \hat{H}_\eff^{(M)}\ket{J_n^{(M)}} = E_n^{(M)}\ket{J_n^{(M)}}.
\end{equation}
Note that the $M$-th order BCH Hamiltonian $\hat{H}_\eff^{(M)}$ is one dimensional and time independent,
which means that the classical system associated with $H_\mathrm{cl}^{(M)}$ is completely integrable.

Now let
\begin{equation}\label{eq:qres}
E^{(\mathrm{res}, n)}_{k} = E_n + k\hbar \Omega,\quad k \in \mathbb{Z},
\end{equation}
be single or multi photon (quantum) absorption energies, {i.e.},  the energy excited by the periodic perturbation with the angular frequency $\Omega$. 
We say that the eigenenergy $E_m$ ($m \neq n$) is a resonance energy with respect to $E_n$ and the quantum resonanceoccurs between the states $\ket{J_n}$  and $\ket{J_m}$ if the condition $E_m = E^{(\mathrm{res}, n)}_{k}$ holds for some $k$.
Correspondingly, the quasi energies $\E_n$ and $\E_m$ create an avoided crossing 
associated with the quantum resonance between $E_n$ and $E_m$. 
Note that here we assume that the symmetry of $\ket{J_m}$ agrees with that of $\ket{J_n}$.

\subsection{\uppercase{Symmetries}}

\begin{figure*}
  \centering
  \includegraphics[width=0.47\textwidth]{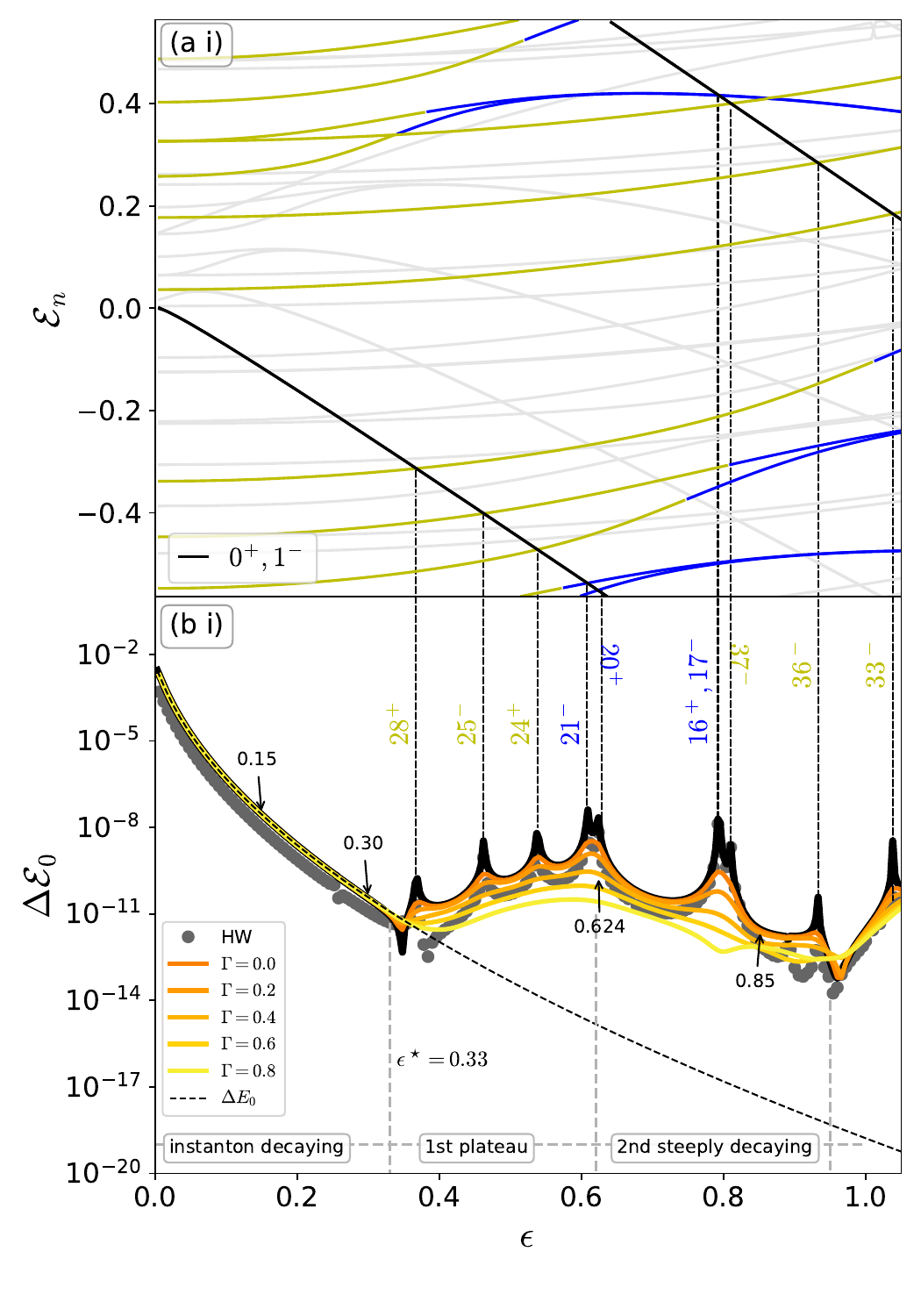}
  \includegraphics[width=0.47\textwidth]{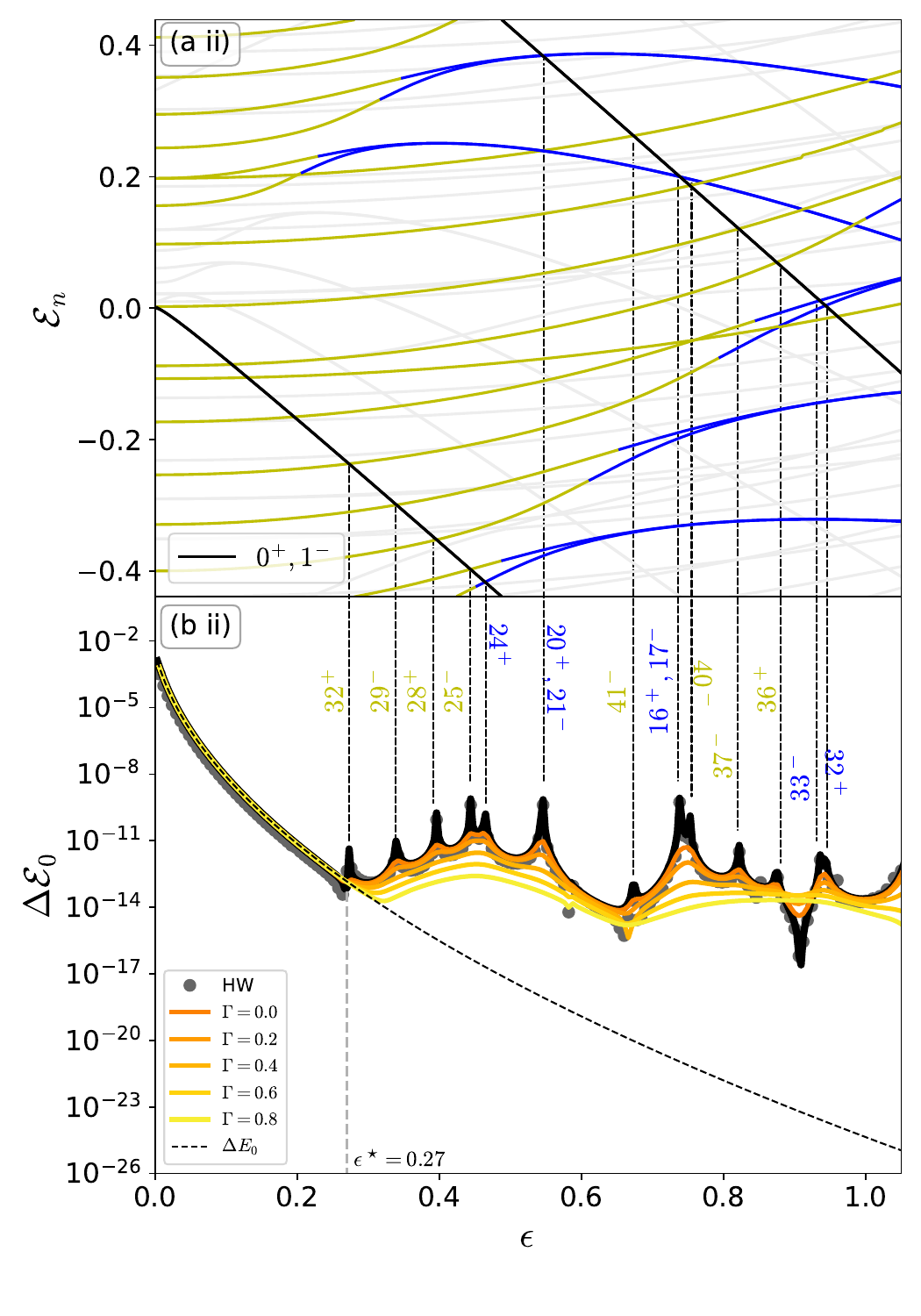}
  \caption{\label{fig:spl_vs_eps}
    (a) Plot of $\E_n$ vs $\epsilon$.
    The solid curves show the quasi energy spectrum for 
    (i) $\hbar=4\pi/70$ and (ii) $4\pi/90$.
    The black curve shows the energies of the ground-state doublet.
    The yellow and blue lines represent the levels for $E_n > \epsilon$ and $E_n<\epsilon$, respectively.
    (b) Plot of $\Delta \E_0$ vs $\epsilon$. The black curve shows the tunneling splitting $\Delta \E_0$ for the ground-state doublet. 
    The dashed curve shows the tunneling splitting of the ground-state doublet for the BCH Hamiltonian.
    The gray dots represent the tunneling splitting obtained by the Herring-Wilkinson splitting formula (\ref{eq:herring}).    
    Quantum numbers of the third (resonant) states are inserted in the figures together with $+$'s or $-$'s.
    The $+$ and $-$ states interact with $\Psi_0^{(+,+)}$ and $\Psi_1^{(-,-)}$ to create the spikes.
    Orange-yellow curves indicate the absorbed splitting $\Delta \Eo_0$ for the ground-state doublet. 
    The lists $L$ of the absorber are (b i) $L=\{28,25,24,21,20,16,17,37,36,33\}$ and (b ii) $L=\{32, 29,28,25,24,20,21,41,16,17,40,37,36,33,32\}$.
    The absorbing perturbation is applied over the entire $\epsilon$ regime.
  }
\end{figure*}

In order to discuss tunneling splittings,
we impose the periodic boundary condition on the phase-space as $(q,p)= (-2\pi,2 \pi ]\times(-\pi,\pi]$.
The number of states up to the energy $H_\mathrm{cl}(p,q) = \eps$ corresponding to the separatrix 
is evaluated using the formula
\begin{equation}
\nsep = \left\lfloor \frac{\A_\eps}{2\pi\hbar} - \frac{1}{2} \right\rfloor.
\end{equation}
Here $\A_\eps$ denotes the area enclosed by the separatrix 
given by the condition 
$H_\mathrm{cl} = \epsilon$
[see Figs.~\ref{fig:stan_ps}(a) and 1(b)].

We note the symmetry of the wave function.
In the $q$ representation,
the eigenstate $J_n(q):=\bracket{q}{J_n}$ has two symmetries:
parity symmetry $J_n(-q)=\pm J_n(q)$ and translational symmetry $J_n(q+2\pi)=\pm J_n(q)$.
The quasi eigenstate $\Psi_n(q):=\bracket{q}{\Psi_n}$ of the quantum map 
has the same symmetries as the integrable one.
We will employ the notation $(\pm,\pm)$ in order to specify the parity and translational symmetries, respectively.
For the librational modes ($n < \nsep$), tunneling splittings are created between a pair
of the states:
\begin{equation}
 \begin{split}\label{eq:spl_rot}
  &\Delta E_n = E^{(-,-)}_{2n+1} - E^{(+,+)}_{2n} \quad \text{for even $n$},\\
  &\Delta E_n = E^{(+,-)}_{2n+1} - E^{(-,+)}_{2n} \quad \text{for odd $n$}.
  \end{split}
\end{equation}
In the cosine-potential case,
the rotational modes ($n>\nsep$)
provide tunneling splittings between a pair of states given as
\begin{equation}
\begin{split}\label{eq:spl_lib}
  &\Delta E_n = E^{(-,+)}_{2n} - E^{(+,+)}_{2n-1} \quad \text{for even $n$},\\
  &\Delta E_n = E^{(-,-)}_{2n} - E^{(+,-)}_{2n-1} \quad \text{for odd $n$}.
\end{split}
\end{equation}

In Fig.~\ref{fig:spl_vs_eps} (a), we plot the energy spectra as a function of $\epsilon$. 
The yellow and blue lines represent the levels for $E_n > \epsilon$ and $E_n<\epsilon$, respectively. 
The levels with symmetries $(+,-)$ and $(-,+)$ are grayed out, since we are focusing on the $n = 0$ doublet (here called the ground-state doublet).
We can see that some yellow lines change their color to blue, 
and then they approach the other blue lines.
We notice that some levels forming the doublets within the librational mode 
change their partner state to the states belonging to different symmetry classes 
when they pass through the separatrix [see Eqs.~(\ref{eq:spl_rot}) and (\ref{eq:spl_lib})]. 
This implies that doublets are not observed in the vicinity of the separatrix 
since the doublets are in the process of changing their partners. 
In the following, we will omit the indices specifying the symmetries, unless it causes confusion.

Numerical calculations are performed by using arbitrary precision arithmetic in {\it Mathematica}, PYTHON with the {MPMATH} package, and {MATLAB} with the {ADVANPIX} toolbox~\cite{mct2015}.

\section{tunneling splitting $\Delta \E_n$ vs $\epsilon$}
\label{sec:spl_vs_eps}

Figure \ref{fig:spl_vs_eps} displays the quasi energy spectrum $\E_n$ as a function of $\eps$ for several values of $\hbar$,
together with the tunneling splitting $\Delta \E_n$ (black solid curve) and $\Delta E_n$ (black dashed curve).
As can be seen, the spectrum has a periodic structure in $\E \in (-\pi\hbar,\pi\hbar]$ reflecting the periodic perturbation in time.
The tunneling splitting in the non-integrable system switches from a smooth to a non smooth dependence,
whose switching point is denoted by $\eps^\star$ in Fig. \ref{fig:spl_vs_eps}(b).

In the regime $\eps > \eps^\star $, the tunneling splitting $\Delta \E_n$ [black solid curves in Figs.~\ref{fig:spl_vs_eps}(b)] shows the non-smooth behavior accompanied by spikes, which persistently deviate from the integrable one, which is shown by a dashed curve in each figure.
Avoided crossings with a third state cause spikes, and we specifically refer to such spikes as quantum resonance (\ref{eq:qres}).

The quantum numbers of the third states are shown with a guideline whose number with the symbols $+$ and $-$ implies the quantum number of the third state.
The label $+$ or $-$ means that the third state has the same symmetry as $\ket{\Psi_0^{(+,+)}}$ or $\ket{\Psi_1^{(-,-)}}$, respectively.
The yellow and blue curves in Fig.~\ref{fig:spl_vs_eps}(a) represent the rotational and librational modes, respectively, that
interact with the ground-state doublet to form avoided crossings in the parameter regime.

If the third state is supported by a chaotic region in the phase-space, we may assume that the resulting spikes are caused by the interaction with chaotic states. 
The enhancement of tunneling splitting induced by such a mechanism is called chaos-assisted tunneling~\cite{bohigas1993,bohigas1993a,tomsovic1994}.
If chaotic regions are significantly developed in the phase-space, 
avoided crossings associated with chaotic states are expected to occur frequently. 
Such a signature could be seen as a manifestation of classical chaos in tunneling splittings.
In the nearly integrable regime,
however, the area of chaotic regions is not large enough to identify chaotic states.
Therefore, we will not use the term chaotic states in the following.

References~\cite{ozorio1984,heller1995,bonci1998,wisniacki2011} pointed out that the classical nonlinear resonances
(the so-called Poincar{\'e}-Birkhoff chains) can also induce avoided crossings.
On the basis of this observation, Ref.~\cite{brodier2002} developed a hybrid method to evaluate the tunneling splitting $\Delta \E_n$
based on the quantum perturbation treatment whose coupling strength is determined by the classical phase-space information.
The mechanism implied by this recipe is called the resonance-assisted tunneling.
Further improvements have been made in \cite{lock2010,schlagheck2011}.

In the following argument, we would like to draw the reader's attention to a difference between a single spike,
which occurs as a result of local interaction between the doublet and a third state,
and the persistent enhancement of tunneling splittings typically observed in the $\Delta \E_n$~vs~$\eps$ plot~\cite{bohigas1993,bohigas1993a,tomsovic1994,bonci1998,wisniacki2011},
or the $\Delta \E_n$~vs~$1/\hbar$ plot~\cite{roncaglia1994,brodier2002,eltschka2005,mouchet2006,mouchet2007,lock2010,deunff2010,backer2010,deunff2013,mertig2013,mertig2016,wisniacki2015}.
So far, the spikes and the persistent enhancement were not clearly distinguished. 
However, we will present evidence that they are phenomena with different origins.

\subsection{\uppercase{absorbing perturbation}}\label{sec:abs}

In this section we introduce the absorbing perturbation, which allows us to 
suppress the interaction with third states and observe its influence on the tunneling splitting $\Delta \E_n$.
To this end, we first formulate the absorbing perturbation method~\cite{hanada2015}.

Let $\ket{J_n}$ be the basis of the $M$th-order BCH Hamiltonian $H_\eff$.
We introduce the absorbing operator
\begin{equation}
  \hat{P} = \1 - \frac{\Gamma}{2}\sum_{\ell\in L} \ketbra{J_\ell}{J_\ell},
\end{equation}
where $\Gamma$ is the strength of the absorbing perturbation and $L$ is the list of quantum numbers that specify the states to be absorbed.
Then we consider the (right) eigenvalue equation for the absorbing operator
\footnote{The symmetric form $\hat{P}\hat{U}\hat{P}^\ast$ may be better to keep the original symmetry.},
\begin{equation}
  \hat{P}\hat{U}\ket{\psio_n} = \uo_n \ket{\psio_n},\qquad \uo_n = e^{-\frac{i}{\hbar} \Eo_n},
\end{equation}
where the absorbed quasi-energy $\Eo_n$ takes a complex value in general.
Note that the quantum number is assigned in the same manner as the closed system.

Assuming $\Gamma$ is a small parameter,
we can apply the second-order perturbation leading to
\begin{equation}\label{eq:pert_eval}
\uo_n \simeq u_n\Bra{1 - \frac{\Gamma}{2}\sum_{\ell\in L} |a_{\ell,n}|^2
  + \frac{\Gamma^2}{4}\sum_{\ell\in L }\sum_{m\neq n} \frac{|a^{\ast}_{\ell,n}a_{\ell,m}|}{u_n-u_m}u_m \cdots
},
\end{equation}
where
\begin{equation}
a_{\ell,n} = \bracket{J_\ell}{\Psi_n}.
\end{equation}
Here the asterisk stands for the complex conjugate.
Since the second term in Eq.~(\ref{eq:pert_eval}) is real, it controls the decay rate.
The first-order absorbed (right) quasi eigenstate is expressed as
\begin{align}
\ket{\psio_n} & \simeq \ket{\Psi_n}- \frac{\Gamma}{2}\sum_{m\neq n} b_{m,n}\ket{\Psi_m},\\
& = \ket{\Psi_n} - \frac{\Gamma}{2} \sum_{m\neq n}\sum_{k} b_{m,n}a_{k,m}\ket{J_k},
\end{align}
where
\begin{equation}
  b_{m,n} = \sum_{\ell\in L}\frac{a_{\ell,m}^\ast a_{\ell,n}}{u_n -u_m} u_m.
\end{equation}
Quantum perturbation theory gives us an intuitive interpretation for the absorbing perturbation,
i.e., the absorbing operator subtracts $\ket{J_k}$ from the exact state $\ket{\Psi_n}$.
Therefore, it can be considered a subtractive perturbation compared with the standard additive perturbation.
Furthermore, the subtraction weight becomes larger as 
the system approaches the resonance condition, i.e., $u_m$ and $u_n$ get closer to each other.

Under this setting, we define the absorbed quasi-energy splitting as
\begin{equation}
\Delta \Eo_0 = |\Eo_{1} - \Eo_{0}|.
\end{equation}
Note that the absorbed quasi-eigenenergy $\Eo_n$ is complex, and so $\Delta \Eo_n$ will be evaluated as a distance in the complex plane.
Thus if $\mathrm{Im} \Delta \Eo_n$ becomes larger than $\mathrm{Re} \Delta \Eo_n$, then $\Delta \Eo_n$ becomes greater than $\Delta \E_n$,
which means that the decay process induced by the absorber overwhelms the tunneling oscillation between the wells.
Therefore, we have to introduce the absorbers carefully so as not to destroy the original spectrum.

\subsection{\uppercase{results}}

The green-yellow curves in Fig.~\ref{fig:spl_vs_eps}(b) represent the absorbed tunneling splitting for the ground-state doublet $\Delta \Eo_n$ with different values of $\Gamma$.
The list $L$ associated with the absorbing perturbation is presented in the figure caption.

As shown in Fig.~\ref{fig:spl_vs_eps}(b), the absorbing perturbation suppresses the spikes,
which means that the influence of the third (resonant) states is removed from the tunneling splitting $\Delta \E_n$.
Nevertheless, the splitting $\Delta \Eo_n$ still keeps deviating from the integrable one.
Moreover, the absorbing perturbation reveals the staircase like structure of the splitting $\Delta \Eo_n$,
which 
is formed by the repetition of plateaus and steeply decaying parts.

This result strongly implies that the effect of interaction with a third state is only around the spike.  
In other words, the range of interaction creating spikes is limited to a rather short range. 
It should be noted that the deviation from integrable tunneling nevertheless remains even 
after removing spikes. 
A similar result was obtained by applying a weak Wick rotation ($\theta \ll 1$) to a quantum map~\cite{mouchet2007}.

\section{The wave function-based observation for the tunneling splitting $\Delta \E_n$}
\label{sec:wave_function_based}

\begin{figure}[t]
	\centering
  \includegraphics[width=0.5\textwidth]{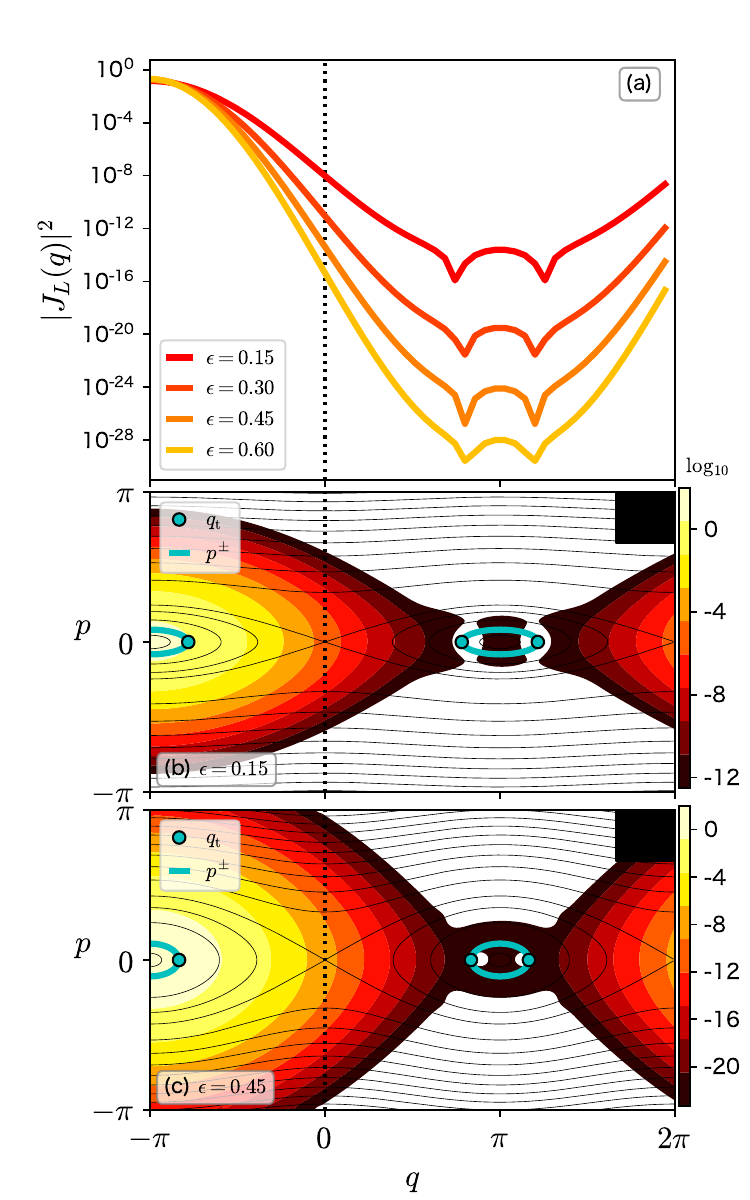}
  \caption{\label{fig:hsm-bch}
    (a) Wave functions $J_L(q)$ for the ground-state doublet of the BCH Hamiltonian
    with $\hbar=4\pi/70$.
    (b) and (c) Husimi representation
    of $J_L(q)$ in $\log_{10}$ scale with several different values of $\epsilon$.
    The associated manifolds $p^{\pm}(q; E)$ and the turning points $q_\mathrm{t}$ are shown by the cyan curves and the cyan dots, respectively. 
    The black box in (b) and (c) shows the effective Plank cell.
  }
\end{figure}

In the regime $\eps < \eps^\star$, as shown in Fig.~\ref{fig:spl_vs_eps},
the tunneling splitting $\Delta \E_n$ decays monotonically 
as a function of $\eps$ and can be well approximated by the splitting for the integrable limit.
From this result, one might think that the mechanism of tunneling in this regime is essentially the same as that of the integrable system in this regime, and can be well captured by it~\cite{backer2010,lock2010,mertig2016}. 
In this section we will show that even in the regime $\eps < \eps^\star$ tunneling tails exhibit  signatures that are different from those appearing in the completely integrable system. 
This is done by performing a wave-function-based observation via the Herring-Wilkinson formula. 
At the same time, our analysis explains why, as explained in Sec. \ref{sec:spl_vs_eps}, the integrable approximation succeeded in reproducing the behavior of tunneling splitting, even though the system is nonintegrable.

The $\eps$ dependence of the tunneling splitting $\Delta E_n$ for the integrable system, e.g.,
$H_1(p,q)=p^2/2 + \epsilon\cos q$, can be obtained by applying the WKB calculation:
\begin{equation}\label{eq:spl_scl}
\Delta E \simeq \frac{\omega_\eps(E)}{\pi \hbar} e^{i S_\epsilon/\hbar}, 
\end{equation}
where $S_\epsilon=\int_{-q_\mathrm{t}}^{q_\mathrm{t}}p(q;E)dq.$
Here the turning point $q_\mathrm{t}$ is defined by the zeros of $p(q;E)$, and $\omega_\eps(E)$ is the classical angular frequency.
Note that there is a pair of solutions $\pm p(q;E)$.
One of the solutions with $\mathrm{Im} p(q; E) < 0$, leading to a divergent contribution,
is to be removed as a result of the Stokes phenomenon, while the other branch gives a tunneling contribution. 
The latter orbit, running in the purely imaginary-time direction, is often referred to as the instanton~\cite{coleman1988,schulman1996}.
Therefore we call the first decaying parameter region the instanton region [see the inset in Fig.~\ref{fig:spl_vs_eps}(b i)]. 
It is also referred to as direct coupling~\cite{tomsovic1994} or direct tunneling~\cite{brodier2002,backer2010,lock2010}. 

For $\hbar \ll 1$, 
the energy of the ground-state doublet can be approximated as $E=-\eps$, and the complex (instanton) action is evaluated as 
\begin{equation}\label{eq:inst}
S_\eps=8\sqrt{-\eps},
\end{equation}
in the integrable limit~\cite{mouchet2007}.
Thus the tunneling splitting $\Delta E_n$ decreases monotonically as a function of $\epsilon$.

\subsection{\uppercase{the Herring-Wilkinson splitting formula}}

The Herring-Wilkinson splitting formula~\cite{herring1962,wilkinson1986,creagh1998} 
is a formula that allows us to evaluate the tunneling splitting in terms of the 
wave function that forms a tunneling doublet. 

Let us consider a one-dimensional time-independent Hamiltonian $H=p^2/2 + \eps V(q)$
with a symmetric double well potential.
Let $\psi_L(q)$ and $\psi_R(q)$ 
be the states associated with congruent equienergy manifolds located in the left and right regions, respectively.
Assume further that the exact eigenstates can be approximated as 
\begin{equation}
\Psi^\pm(q) \simeq \frac{1}{\sqrt{2}}\BRa{\psi_L(q) \pm \psi_R(q)},
\end{equation}
with energies $E\pm\Delta E/2$. 
For the sake of simplicity, we will drop the quantum number $n$ here.
Due to the (one-dimensional and time-independent) Herring-Wilkinson splitting formula, the tunneling splitting between two states $\psi_L(q)$ and $\psi_R(q)$ is evaluated as
\begin{equation}\label{eq:herring}
  \Delta E \simeq \hbar^2\Bra{\psi'^\ast_R(q) \psi_L(q) - \psi_R^\ast(q) \psi'_L(q)}_{q=0},
\end{equation}
where the prime stands for the derivative with respect to $q$.
Assuming further that $\psi_{L(R)}(q)$ is expressed by a local WKB solution around $q=0$,
\begin{align}\label{eq:local-wkb}
\psi_{L(R)} \sim e^{\frac{i}{\hbar} S_\eps(q;E)}, \quad S_\eps(q;E)=\pm \int_{\mp q_\mathrm{t}}^{q}p(q';E) dq',
\end{align}
and inserting (\ref{eq:local-wkb}) into the formula (\ref{eq:herring}),
we reach the semiclassical formula of the tunneling splitting (\ref{eq:spl_scl}).
Here $q_\mathrm{t}$ denotes the turning point close to $q=0$.

In nearly integrable systems, according to the Kolmogorov-Arnold-Moser (KAM) theorem, 
invariant curves supporting the WKB states survive under small perturbations. 
However, one cannot develop WKB theory for tunneling splittings in nearly integrable systems in the same way because KAM curves do not bridge congruent equienergy invariant manifolds 
in the complex plane, but are expected to have a border of analyticity~\cite{greene1981,berretti1990,berretti1992}.
This prevents us from constructing semiclassical states $\psi_{L(R)}(q)$. 
Therefore, to compute wave functions based on the Herring-Wilkinson splitting formula, 
we use here localized wave functions constructed numerically, instead of the WKB states,
\begin{equation}\label{eq:LR}
\Psi_{L(R)}(q) = \frac{1}{\sqrt{2}}\BRa{\Psi_0(q) \pm \Psi_1(q)},
\end{equation}
where $\Psi_n(q)$ $(n=0,1)$ are states obtained by direct numerical calculations.

The Herring-Wilkinson splitting formula for Floquet systems can be applied analogously to the time-independent one
(see the Appendix \ref{app:spl}).
The gray dots in Figs.~\ref{fig:spl_vs_eps}(b) represent the tunneling splittings calculated based on the Herring-Wilkinson splitting formula.
The derivatives in Eq.~(\ref{eq:herring}) are evaluated numerically using the send-order (central) difference scheme.
As can be seen in Fig.~\ref{fig:spl_vs_eps}, the Herring-Wilkinson splitting formula works well, 
suggesting that the amplitude $\Psi_{L(R)}(q)$ at $q=0$ controls the tunneling splittings.

\subsection{\uppercase{wave function-based observation}}\label{sec:integ}

\begin{figure*}[t]
  \includegraphics[width=0.245\textwidth,trim={0.75cm 2cm 0 0},clip]{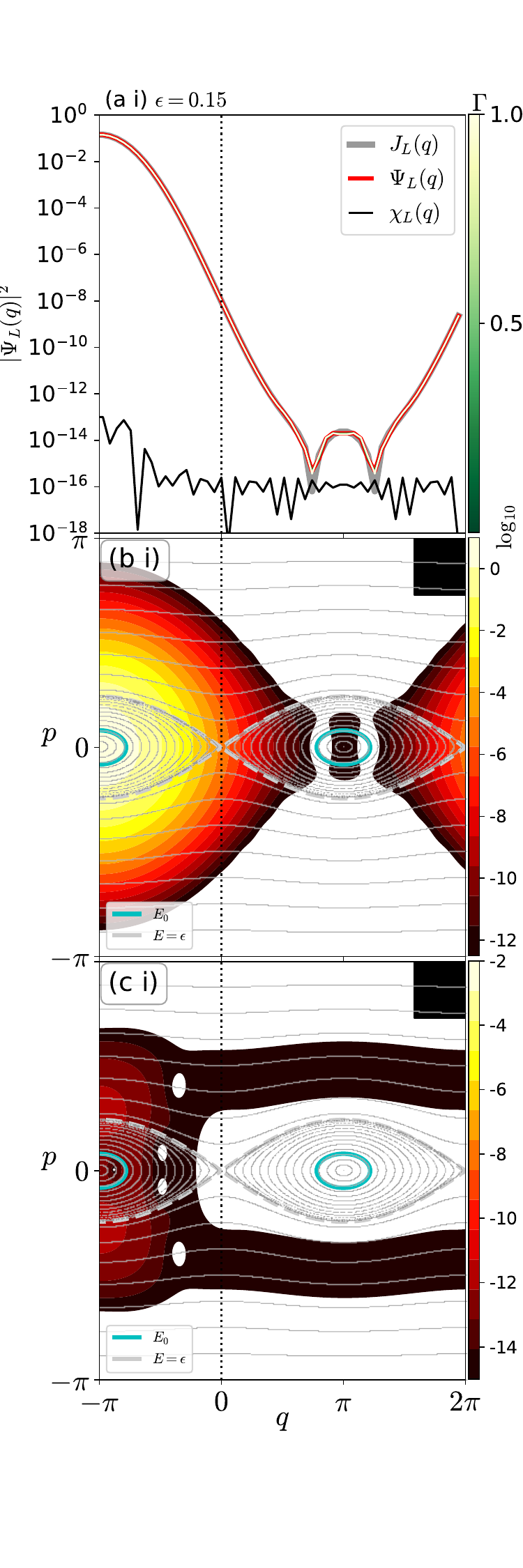}
  \includegraphics[width=0.245\textwidth,trim={0.75cm 2cm 0 0},clip]{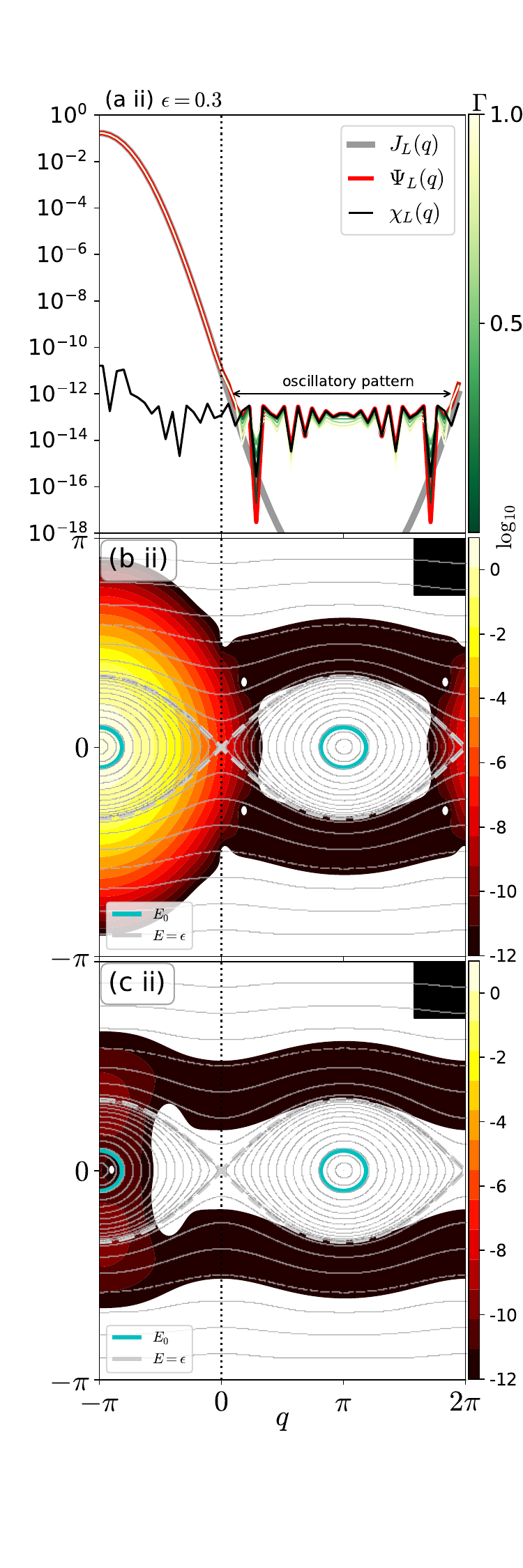}
  \includegraphics[width=0.245\textwidth,trim={0.75cm 2cm 0 0},clip]{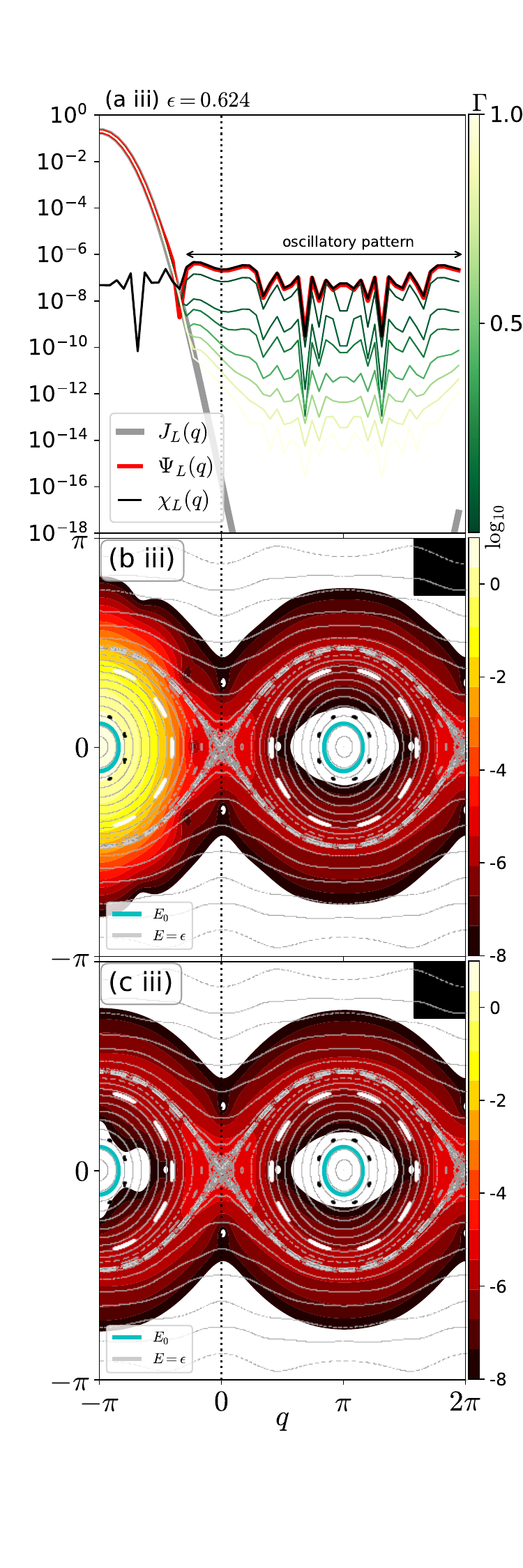}
  \includegraphics[width=0.245\textwidth,trim={0.75cm 2cm 0 0},clip]{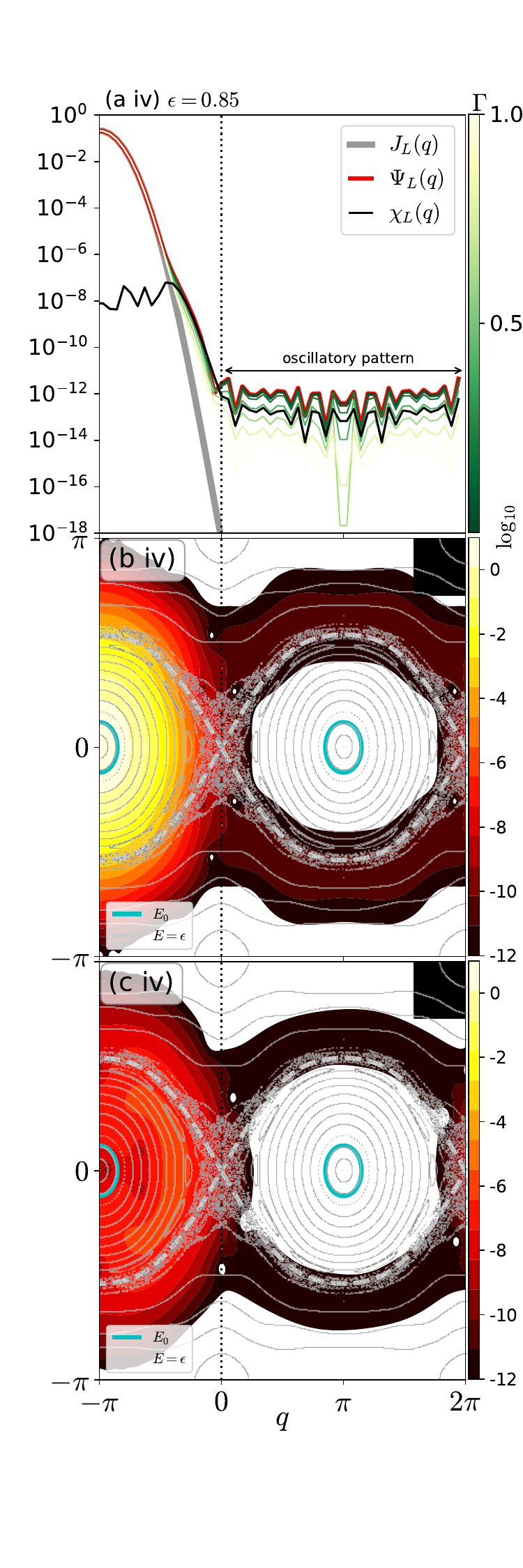}
  \caption{\label{fig:hsm-N70}
    (a)  Wave functions of the quantum map $\Psi_L(q)$ (red), the associated integrable system $J_L(q)$ (gray),
    the perturbation term $\chi_L(q)$ in Eq. (\ref{eq:pert}) (black) ,
    and the absorbed wave function $\psiLo(q)$  (green-yellow curves) whose absorbing perturbation strength 
    $\Gamma$ shown in the color scale to the right. 
    The value of the parameter $\epsilon$, which is also shown in the inset of Fig.~\ref{fig:spl_vs_eps}(b i),
    is (a i) 0.15, (a ii) 0.3, (a iii) 0.624 and (a iv) 0.85.
    The Husimi representations of (b) $\Psi_L(q)$ and (c) $\chi_L(q)$ are shown in $\log_{10}$ scale.
    The cyan solid line and the light gray dashed curve in (b) and (c) represent the contour curves specified by $H_\mathrm{cl}=E_0$ and $H_\mathrm{cl}=\eps$, respectively.
    The Planck constant is chosen as $\hbar=4\pi/70$. The black boxes in (b) and (c) stand for the effective Planck cell. }
\end{figure*}

Before moving on to the analysis for the non-integrable system,
we will discuss integrable systems for reference.
Let $\ket{J_0}$ and $\ket{J_1}$ form the ground state doublet for the BCH basis. 
The localized states are constructed as
\begin{equation}
J_{L(R)}(q) = \frac{1}{\sqrt{2}}[J_0(q) \pm J_1(q)].
\end{equation}
Figure~\ref{fig:hsm-bch}(a) illustrates the wave function $J_L(q)$ for several values of $\eps$.
The WKB argument predicts how wave functions behave as a function of $\epsilon$.
The (local) WKB solution is expressed as Eq. (\ref{eq:local-wkb}) with 
$p(q;E) = \sqrt{2[E - V(q)]}$. 
Since the imaginary action $S_\eps(q)$ increases monotonically with $q$, the resulting WKB wave function decays monotonically and exponentially
from the left turning point $-q_\mathrm{t}$ to the opposite turning point $q_\mathrm{t}$.
Then the manifold supporting the WKB wave function connects to the real branches $\pm p(q; E)$ and generates oscillatory pattern in wave function.
It is difficult to identify the turning points for $q<0$ from the profile of 
$J_L(q)$ because the ground state does not have nodes.
On the other hand, the turning points $q_\mathrm{t}$ for $q>0$ can be 
identified from the wave function $J_L(q)$ since there is a sharp dip at the turning points  $q_\mathrm{t}$. 
Note that the quantum number of the ground state doublet is zero, 
so the oscillatory pattern associated with 
the real branches $\pm p(q; E)$ for $q>0$ appears just as a single convex around $q=\pi$, which is found in Fig. \ref{fig:hsm-bch}(a).

Due to the the Herring-Wilkinson splitting formula~(\ref{eq:herring}), 
one can find that 
the $\eps$ dependence of the tunneling splitting $\Delta E_n$ almost follows 
the $\eps$ dependence of the amplitude of the wave function around $q=0$.
Recall that 
$S_\epsilon(q=0; E)$ is proportional to $\sqrt{-\epsilon}$, so the amplitude of $J_L(q=0)$ decays as $e^{-\sqrt{\epsilon}/\hbar}$ [see Eq. (\ref{eq:inst})]. 
Therefore, the splitting $\Delta E$ decays monotonically as a function of $\epsilon$ or $1/\hbar$.

Now we turn to the non-integrable case.
Figures~\ref{fig:hsm-N70}(a) demonstrates the wave functions $J_L(q)$ (gray curves) and $\Psi_L(q)$ (red curves)
and the absorbed ones $\Psi_L^\Gamma(q)$ (green-yellow curves) for several different values of $\eps$.
The corresponding Husimi representations for $\psi_L(q)$ is shown Figs.~\ref{fig:hsm-N70}(b).

As shown in Fig.~\ref{fig:hsm-N70}(a i), in the region $\eps\ll\eps^\star$, 
the wave function $\Psi_L(q)$ is well approximated by the integrable one $J_L(q)$.
Moreover, there is no noticeable difference in the Husimi-representation between $\ket{\Psi_L}$ and $\ket{J_L}$ [see Fig.~\ref{fig:hsm-N70}(a i) and \ref{fig:hsm-N70}(b i)]. It should also be noted that there are several avoided crossings with the rotational modes in this regime,
but such avoided crossings do not create visible spikes in the tunneling splitting $\Delta \E_n$. 

As $\eps$ approaches $\epsilon^\star$,
on the other hand, although the tunneling splitting $\Delta \E_n$ coincides with the integrable one [see Fig.~\ref{fig:spl_vs_eps}(b i)],
the tunneling tail of $\Psi_L(q)$ significantly deviates from that of $J_L(q)$ 
[see Fig.~\ref{fig:hsm-N70}(a ii)].
Notice that the tunneling tail 
for $\Psi_L(q)$
exhibits oscillatory patterns in the region $0<q<2\pi$,
which originates from the transversal KAM curves running outside of librational KAM curves [see Fig.~\ref{fig:hsm-N70}(b ii)]. 
This observation strongly suggests that tunneling across the separatirx plays an significant role in 
the deviation of $\Psi_L(q)$ from $J_L(q)$. 
According to the Herring-Wilkinson splitting formula, we can predict the magnitude of the tunneling splitting from the amplitude of the wave function at $q=0$.
Indeed, we can see in Fig.~\ref{fig:hsm-N70}(b ii) that the amplitude of $\Psi_L(q)$ at $q=0$ matches that of $J_L(q)$.

For the regime $\eps>\eps^\star$, on the other hand,
the oscillatory pattern spreads across the separatrix and covers the region containing $q=0$
[see Fig.~\ref{fig:hsm-N70}(a iii)].
As a result, the tunneling splitting $\Delta \E_n$ no longer follows the integrable one and starts to deviate [see Fig.~\ref{fig:spl_vs_eps}(b i)].
We also recognize that the wave function $\Psi^\Gamma_n(q)$ keeps an oscillatory pattern 
even when the absorbing perturbation is applied. 
In particular, the amplitude at $q=0$ is several orders of magnitude larger 
than that in the integrable one. 
This robustness gives rise to the persistent deviation of the tunneling splitting $\Delta \E_n$
from the integrable prediction, 
meaning that once the deviation appears, the tunneling splitting $\Delta \E_n$ 
can never be approximated by the integrable splitting. 
The existence of an oscillatory pattern implies that multiple modes are involved in creating the wave function 
(see Fig.~\ref{fig:hsm-N70}).

As shown here, the wave function is more informative than the tunnel splitting. 
The tunnel splitting concerns only the value of the wave function at $q=0$, but the wave function reveals how the deviation from the integrable tunneling proceeds. In particular, even in the parameter regime where the tunnel splitting obtained 
from the integrable approximation well approximates that for the nonintegrable map, 
we can detect the difference in the wave function. 

\subsection{\uppercase{tunneling across separatrix}}\label{subsec:pert}

\begin{figure}[t]
	\centering
	\includegraphics[width=0.5\textwidth]{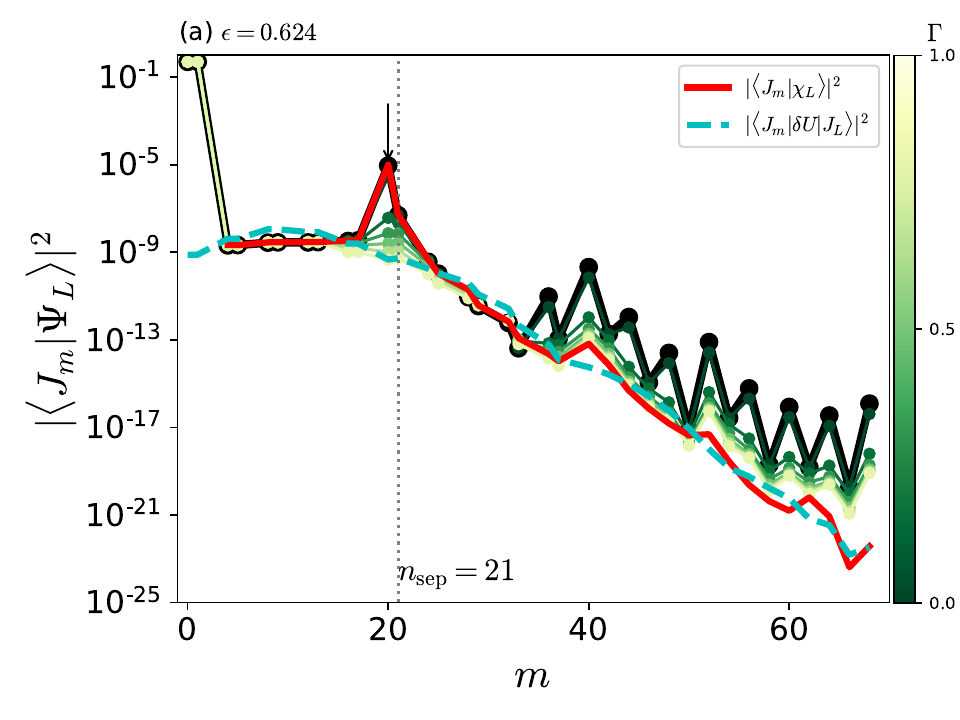}
	\includegraphics[width=0.5\textwidth]{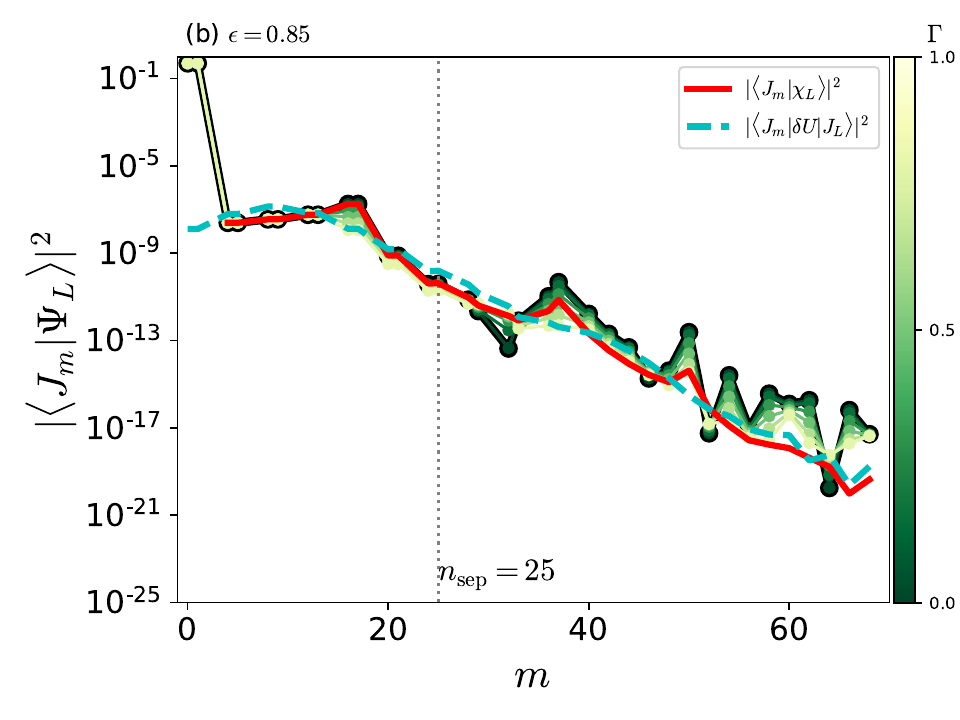}
	\caption{\label{fig:matele_vs_eps}
		(color online)
   Projection of the exact wave function $\ket{\Psi_L}$ (black solid curves) and the absorbed one $\ket{\psiLo}$
   (green-yellow curves) onto the basis states $\ket{J_m}$ for (a) $\epsilon=0.624$ and (b) $\epsilon=0.85$.
	  Cyan dashed and red solid curves represent $\bra{J_m}\delta U\ket{J_L}$ and $\bracket{J_m}{\chi_L}$, respectively.
		The black arrow in (a) displays the resonant (third) state [see Fig.~\ref{fig:spl_vs_eps}(a~i) and \ref{fig:spl_vs_eps}(b~i)].
	}
\end{figure}

As discussed in Refs.~\cite{shudo2014,hanada2015},
the quantum perturbation calculation based on the BCH Hamiltonian reproduces the tunneling tail of the exact eigenstates.
In this section  we further proceed with wave-function-based observations using quantum perturbation theory, and examine the nature of the oscillatory patterns that are thought to be caused by the tunneling across the separatrix, found in the preceding section  (see Fig.~\ref{fig:hsm-N70}).

The one-step time evolution generated by the $M$th-order BCH Hamiltonian
$\hat{U}^{(M)}_\eff:=e^{-\frac{i}{\hbar}\hat{H}_\eff^{(M)}}$
is very close to the time evolution driven by the quantum map $\hat{U}$.
Now we introduce the residual operator as
\begin{equation}
\delta \hat{U}_M = \hat{U} - \hat{U}^{(M)}_\eff,
\end{equation}
which can be made sufficiently small 
as we increase the truncation order $M$. 
The residual operator $\delta \hat{U}_M$ can be regarded as the small perturbation for the integrable time-one evolution $\hat{U}_\mathrm{eff}^{(M)}$.
For 
the optimal truncation $M$,
the exact state is expanded in terms of BCH eigenstates as a perturbative form:
\begin{equation}\label{eq:pert}
\ket{\Psi_n}\approx \ket{\Psi_n^{(\mathtt{BCH})}}=\ket{J_n} + \ket{\chi_n},
\end{equation}
where
\begin{equation}\label{eq:chi}
\ket{\chi_n} := \sum_{m\neq n} \frac{\bra{J_m}\delta \hat{U}_M\ket{J_n}}{e^{-\frac{i}{\hbar}E_m } - e^{-\frac{i}{\hbar}E_n }}\ket{J_m}.
\end{equation}
Note that the standard perturbation theory consists of the known parameters,
while Eq.~(\ref{eq:chi}) contains nontrivial terms related to the residual operator $\delta \hat{U}_M$. 

Using the ground-states doublet calculated by Eq.~(\ref{eq:chi}),
we construct the superposed wave functions $\ket{\chi_{L(R)}} := (\ket{\chi_0} \pm \ket{\chi_1})/\sqrt{2}$.
The black curve in Fig.~\ref{fig:hsm-N70}(a) displays the wave function $\chi_L(q)$,
and the corresponding Husimi representation is shown in Figs.~\ref{fig:hsm-N70}(c).
The parameter values $\epsilon$ plotted in Fig.~\ref{fig:hsm-N70} are marked with arrows in Fig.~\ref{fig:spl_vs_eps}.

For $\eps\ll\eps^\star$, as shown in Fig.~\ref{fig:hsm-N70}(a~i),
the relation $\chi_L(q)< J_L(q)$ holds in the entire $q$ regime,
so the contribution from $\chi_L(q)$ is not visible in the state $\Psi_L(q)$.
As illustrated in Fig.~\ref{fig:hsm-bch}, the tunneling tail of $J_L(q)$ decays exponentially as predicted by the WKB formula (\ref{eq:local-wkb}),
whereas the amplitude of $\chi_{L}(q)$ tends to increase up to the end of the first plateau 
of the $\Delta \E$-$\eps$ plot 
[compare the amplitudes of $\chi_{L}(q)$ (black curves) in Fig.~\ref{fig:hsm-N70}(a i)-\ref{fig:hsm-N70}(a iii)].

As $\epsilon$ increases, the oscillatory pattern mentioned above 
appears and gradually expands to form a plateau. The left edge of the plateau finally 
reaches $q=0$, as seen in Figs.~\ref{fig:hsm-N70}(a ii) and \ref{fig:hsm-N70}(a iii).
Thus, we can specify the critical perturbation strength $\epsilon^\star$ as a parameter value
at which the amplitude at $q=0$ crosses over between $J_L(q)$ and $\chi_L(q)$. 
Figure~\ref{fig:hsm-N70}(a ii) is closest to such a crossover situation. 
This crossover is directly observed in the Husimi representation of $\ket{\chi_L}$.
Note that the oscillatory pattern localized on a transversal KAM curve persists
even in the second steeply decaying region in the $\Delta \E_n$ vs $\eps$ plot [see Figs.~\ref{fig:hsm-N70}(a iv), \ref{fig:hsm-N70}(b iv), and \ref{fig:hsm-N70}(c iv)].

The projection onto the basis $\ket{J_m}$ gives further information on the wave function.
Figure~\ref{fig:matele_vs_eps} displays the amplitudes of the expansion coefficients $\bracket{J_m}{\Psi_L}$ and $\bracket{J_m}{\Psi_L^\Gamma}$ and the transition matrix elements $\bra{J_m}\delta \hat{U}\ket{J_L}$.
The corresponding wave functions in the $q$ representation of Figs.~\ref{fig:matele_vs_eps}(a) and \ref{fig:matele_vs_eps}(b) are drawn in Figs.~\ref{fig:hsm-N70}(a iii) and \ref{fig:hsm-N70}(a iv), respectively.
As shown in Fig.~\ref{fig:matele_vs_eps}(a),
the amplitudes $\bracket{J_m}{\Psi_L}$ and
the perturbed one $\bracket{J_m}{\chi_L}$ show the spikes, marked by the arrow in the plot.
This is because the transition matrix elements do not contain the energy denominator giving rise to resonances whereas $\ket{\chi_L}$ has the energy denominator.

As can be seen in Fig.~\ref{fig:matele_vs_eps}, the amplitude of the resonance peak decreases significantly as the absorbing perturbation strength $\Gamma$ increases.
When $\Gamma$ is close to 1,
the amplitudes $\bracket{J_m}{\psiLo}$ and $\bra{J_m}\delta \hat{U}\ket{J_L}$ 
almost coincide with each other.
This shows that the absorbing perturbation effectively suppresses 
the influence of avoided crossings, 
since the latter amplitude excludes the energy denominator terms in the perturbation.

\begin{figure}[t]
  \centering
  \includegraphics[width=0.5\textwidth]{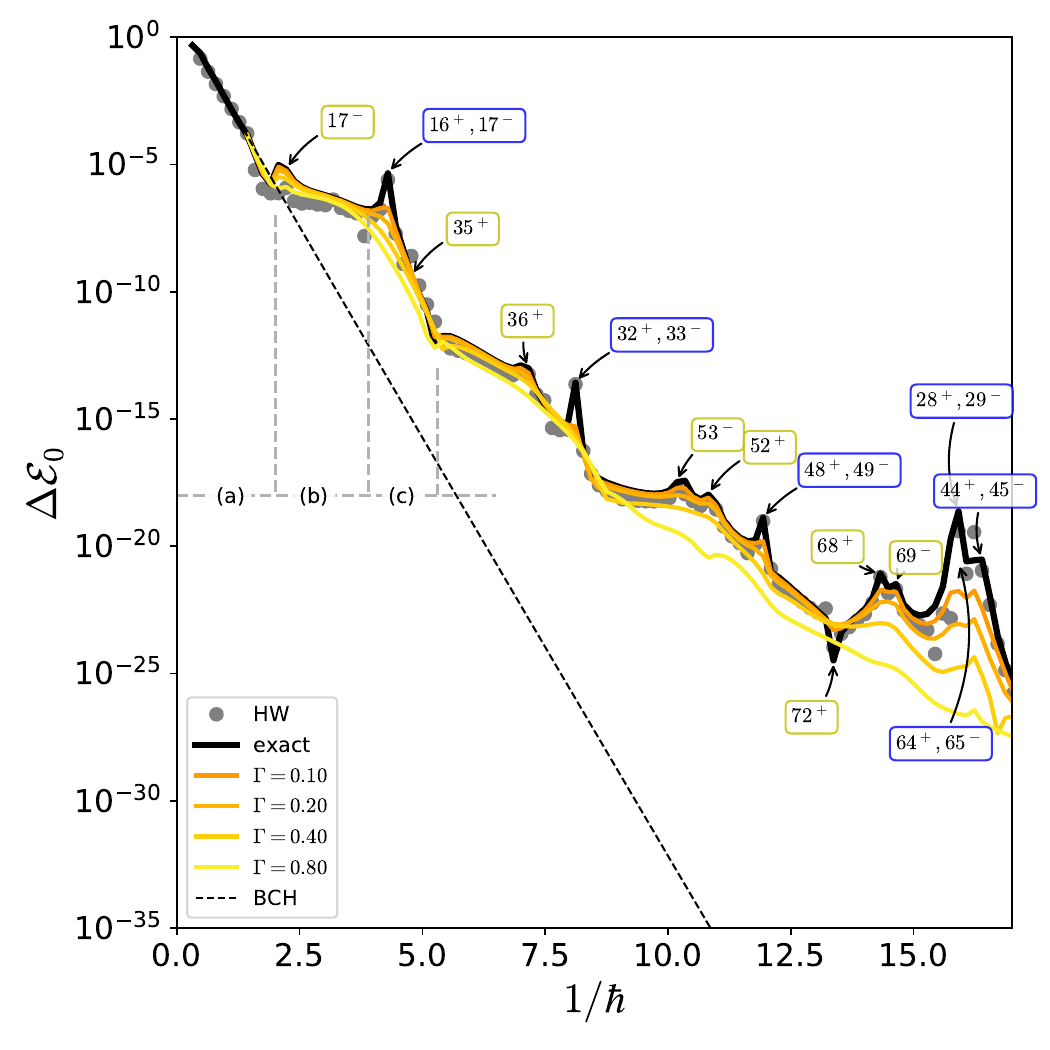}
  \caption{\label{fig:spl_vs_hbar}(color online)
      The black and green-yellow solid curves are the exact tunneling splitting $\Delta \E_0$
      and the absorbed splitting $\Delta \Eo_0$, respectively (the absorbing strength is given in the legend).
      The gray dots represent the tunneling splitting obtained using the Herring-Wilkinson splitting formula.
      The list of quantum states used to construct the absorber is $L=\{ 16,17,18,35,36,37,32,33,52,53,48,49,72,68, 69, 28, 29, 44,\allowbreak 45,\allowbreak 46, 64, 65\}$.
      The black dashed curve represents the tunneling splitting $\Delta E_0$ of the BCH Hamiltonian.
      The numbers in the blue and yellow boxes indicate the quantum number of the resonant (third) states.
      The blue and yellow boxes distinguish whether the third state is a librational or rotational mode, respectively.
      The symbol $+$ indicates if the third state has the same symmetry as that for $\ket{\Psi_0^{(+,+)}}$,
      and the symbol $-$ indicates if it belongs to the opposite symmetry state.
      Regions (a), (b), and (c) refer to
      the first (instanton) decay, the first plateau, and the second steeply decaying region (see the text).
  }
\end{figure}

For even larger $\eps$ shown in Fig.~\ref{fig:matele_vs_eps}(b),
the amplitudes $\bra{J_m}\delta \hat{U}\ket{J_L}$ and $\bracket{J_m}{\Psi_L}$ almost coincide with each other, which means that the influence of 
avoided crossings becomes negligible.
In fact, the absorbing perturbation gives a small effect on the oscillatory pattern of the wave function $\Psi_L^\Gamma(q)$ [see Fig.~\ref{fig:hsm-N70}(a iv)]
and the expansion coefficient for $\bracket{J_m}{\Psi_L^\Gamma}$ [see Fig.~\ref{fig:matele_vs_eps}(b)].
We also present a situation where the value of $\epsilon$ is chosen such that the doublet states are away from any avoided crossings. In this case, quantum resonances do not happen and correspondingly no spikes appear in $\bracket{J_m}{\chi_L }$.

Our observations reveal that
the influence of the 
avoided crossings are well localized around the spikes as discussed in Sec. \ref{sec:abs}.
Furthermore, the plateau region around $q=0$, 
which brings the persistent enhancement of the tunneling splitting $\Delta \E_n$,
is induced by the broadly spreading transition matrix elements  $\bra{J_m}\delta\hat{U}\ket{J_L}$ across the separatrix.
These results provide strong evidence showing that the robustness of the tunneling splitting enhancement, which has been verified in Sec. \ref{sec:persistent}, is due to transition matrix elements $\bra{J_m}\delta\hat{U}\ket{J_L}$ 
with such unique properties.

\section{$1/\hbar$-dependence for the splitting}
\label{sec:spl_vs_hbar}

In the preceding section, we analyzed the behavior of tunneling splittings
as a function of the perturbation strength $\eps$, keeping $\hbar$ fixed.
In this section we fix $\eps$ and observe tunneling splittings as a function of
$1/\hbar$. Such an observation has been made frequently in the literature~\cite{roncaglia1994,brodier2002,eltschka2005,mouchet2006,mouchet2007,schlagheck2011,lock2010,backer2010,mertig2013,mertig2016}.
We show here that the scenario obtained in the preceding section holds true in the latter case as well.

\subsection{\uppercase{Response to the absorbing perturbation}}

\begin{figure}[b]
  \includegraphics[width=0.45\textwidth]{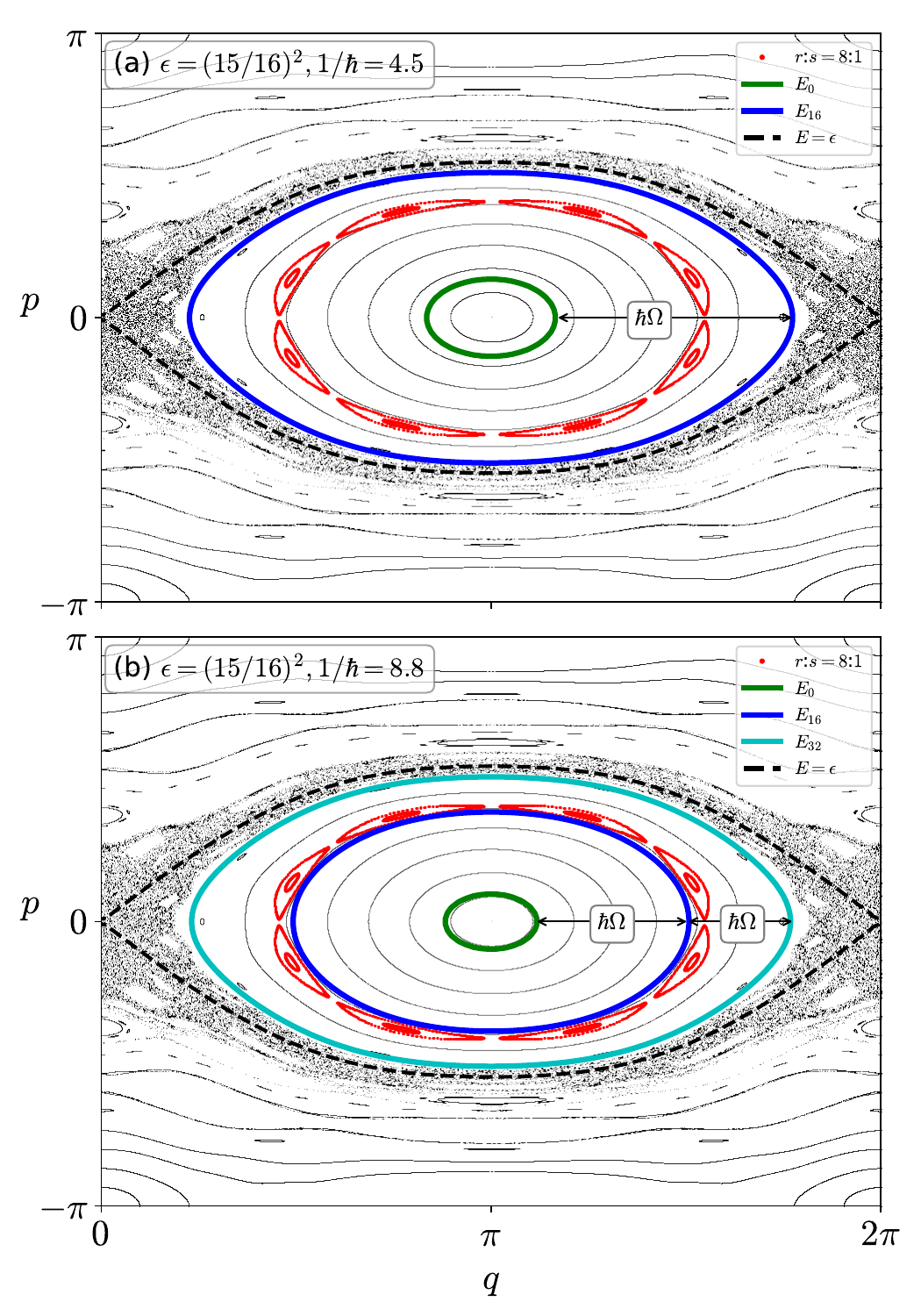}
  \caption{\label{fig:res8_vs_hbar}
     Phase space portraits of the classical map $f$ with $\eps=(15/16)^2$. The classical resonance 
    with $r:s=8:1$ is shown as red dots.
    The energy contours of the BCH Hamiltonian $H_\textrm{cl}$ associated with the RAT scheme for (a) $1/\hbar=4.5$ and (b) $1/\hbar=8.8$ are drawn as solid curves (see the legends).
    The separatrix of the BCH Hamiltonian is shown by the black dashed curve.
  }
\end{figure}

As seen in Fig.~\ref{fig:spl_vs_hbar}, the plot of $\Delta \E_n$ vs $1/\hbar$ typically creates a staircase structure. 
The underlying mechanism for the staircase structure was closely studied in Ref.~\cite{hanada2015}.
In our following argument, we will specifically focus on 
the first steeply decaying part ($1/\hbar<2$), 
which will be called the instanton region hereafter. 
For larger $1/\hbar$, the first plateau ($2<1/\hbar<4$) and 
the second steeply decaying region ($4<1/\hbar<5$) follow 
(see also Fig.~\ref{fig:spl_vs_hbar}).

As is the case in the preceding section, when we apply the absorbing perturbation,
the spikes created as a result of avoided crossings disappear, yet
the staircase backbone remains, as seen in Fig.~\ref{fig:spl_vs_hbar}.
In this calculation, the absorber is made up of the states with 
the quantum number specified in the box of Fig.~\ref{fig:spl_vs_hbar} and applied in the whole $1/\hbar$ regime.
Note that spikes in  the $\Delta \E_n$ vs $1/\hbar$ plot appear around
avoided crossings, 
not necessarily just at a particular avoided crossing point
as in the $\Delta \E_n$ vs $\eps$ plot.

As has been clarified in Ref.~\cite{hanada2015}, and will be discussed below,
the grand-state doublet and excited states are broadly or even almost equally coupled with excited states.
Such a signature explains
the robustness of the staircase structure against the absorbing
perturbation. Note that this scenario is already known from the
analysis for the $\Delta \E_n$ vs $\eps$ plot in the preceding section.

\subsection{\uppercase{Resonance-assisted tunneling and the signature of coupling}}\label{sec:RAT}

In this section we examine whether the scenario revealed through the absorbing perturbation experiment and the coupling nature under the BCH representation could be compatible with the RAT picture by carrying out the calculation 
following the recipe presented in Refs.~\cite{brodier2002,lock2010,schlagheck2011}. 
Within the RAT theory, 
the concrete recipe for calculating quantities related to tunneling via classical nonlinear resonances first makes use of information taken from classical resonances in question and then applies quantum perturbation theory.
In this sense the method is hybrid, combining classical data with the standard quantum perturbation scheme.

\begin{figure}
  \centering
  \includegraphics[width=0.5\textwidth]{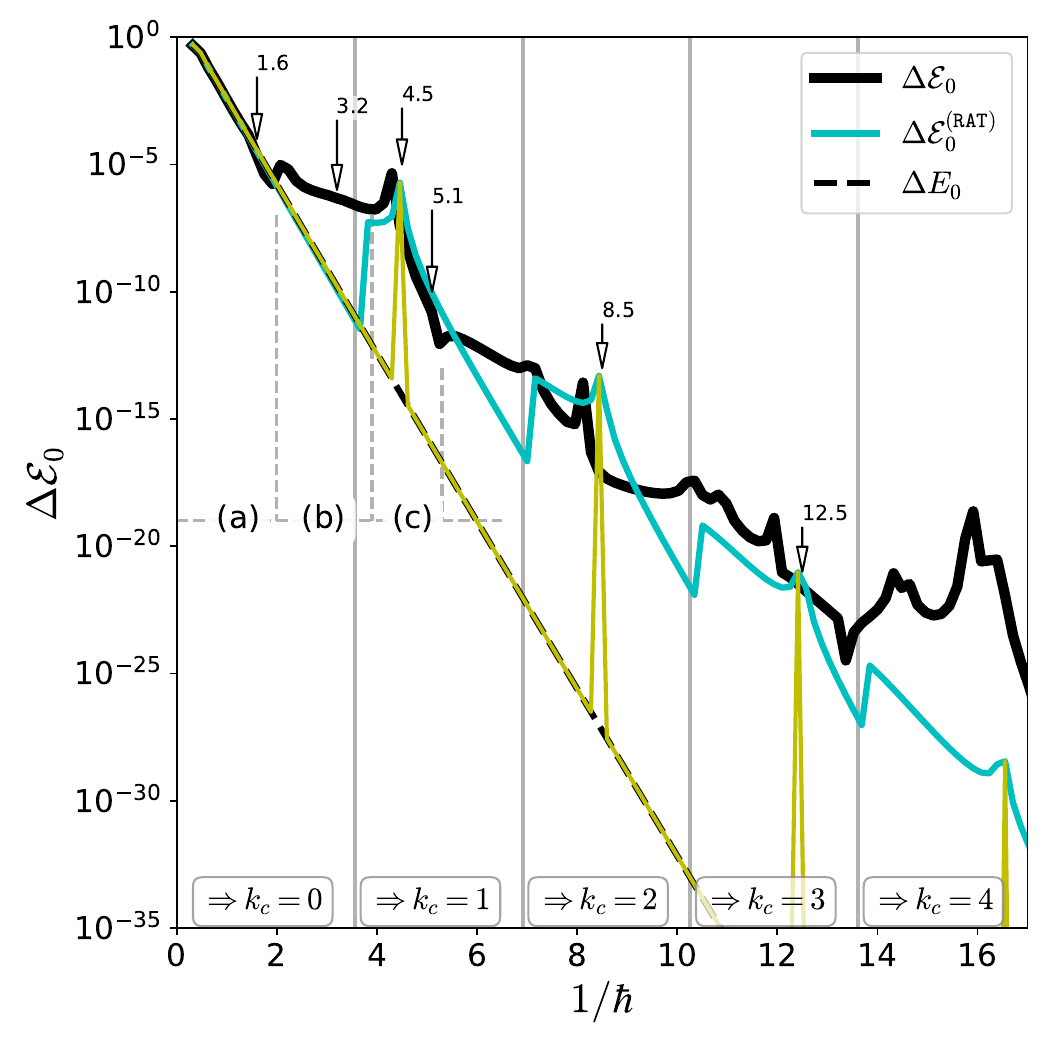}
  \caption{\label{fig:spl_vs_hbar_RAT}
  	The black solid and dashed curves are tunneling splitting $\Delta \E_0$ obtained by the exact calculation 
	and $\Delta E_0$ obtained by diagonalizing the BCH Hamiltonian, respectively.   
	The cyan curve is the tunneling splitting 
	$\Delta \E_0^{(\mathtt{RAT})}$ obtained by the RAT calculation scheme using the $r:s=8:1$ nonlinear classical resonance.
	The gray vertical lines indicate values of $1/\hbar$ at which the value of $k_c$ in Eq.~(\ref{eq:rat_spl}) is incremented by one.
	The $1/\hbar$ regions (a), (b), and (c) are the first steeply decaying (or instanton), first plateau, 
	and second steeply decaying regions (see the text for details).
	The parameter values used for our RAT calculation are
	$S_{r:s}^{+} = 6.573, S_{r:s}^{-}=7.280$, and $\mathrm{Tr}M_{r:s}=1.701\,463\,596\,85$.
  }
\end{figure}

The wave function associated with a classical $r:s$ resonance is locally constructed from the RAT scheme as
\begin{equation}\label{eq:rat_wave}
\ket{\Psi_n^\mathtt{(RAT)}} = \ket{J_n} + \sum_{k>0} B_{n+kr,n} \ket{J_{n+kr}},
\end{equation}
where
\begin{equation}
B_{n+kr,n} = \prod_{\ell=1}^{k_c}  \frac{A_{n+\ell r, n+(\ell-1)r}}{E_n - E_{n+\ell r} + \ell s\hbar \omega},
\end{equation}
with
\begin{equation}\label{eq:rat_coeff}
A_{n+\ell r, n +(\ell-1)r} = V_{r:s}(I_{r:s}) e^{i\phi_k} \Bra{\frac{\hbar}{I_{r:s}}}^{kr} \sqrt{\frac{(n+kr)!}{n!}}.
\end{equation}
Here $E_n$ denotes the unperturbed energy, and
$I_{r:s}$ represents the action satisfying the classical resonance condition $\omega:\Omega=r:s$,
where $\omega$ and $\Omega$ are the internal and external (perturbative) frequencies. 
The matrix element $B_{m,n}$ is sparse and has non zero values only if
 the condition $m = n+kr$ for $k>0$ is satisfied~\cite{brodier2002,schlagheck2011}.
The summation is taken up to
\begin{equation}
k_c = \left\lfloor\frac{1}{r}\Bra{\frac{\mathcal{A_\mathrm{reg}}}{2\pi\hbar} - \frac{1}{2}}\right\rfloor.
\end{equation}
Here $\mathcal{A}_\mathrm{reg}$ stands for the area of the regular tori centered at $(q,p)=(\pm \pi,0)$.
There may be arguments about how to determine the regular region in the corresponding quantum system~\cite{mouchet2006,schlagheck2011,backer2010}. 
Here we assume that the regular region $\mathcal{A}_{\rm reg}$ is the region bounded by 
separatrix $\mathcal{A}_{\epsilon}$ of $H_\mathrm{cl}$ for simplicity.

\begin{figure*}
  \centering
  {
  \includegraphics[width=0.245\textwidth,trim={1.5cm 2cm 0 0}]{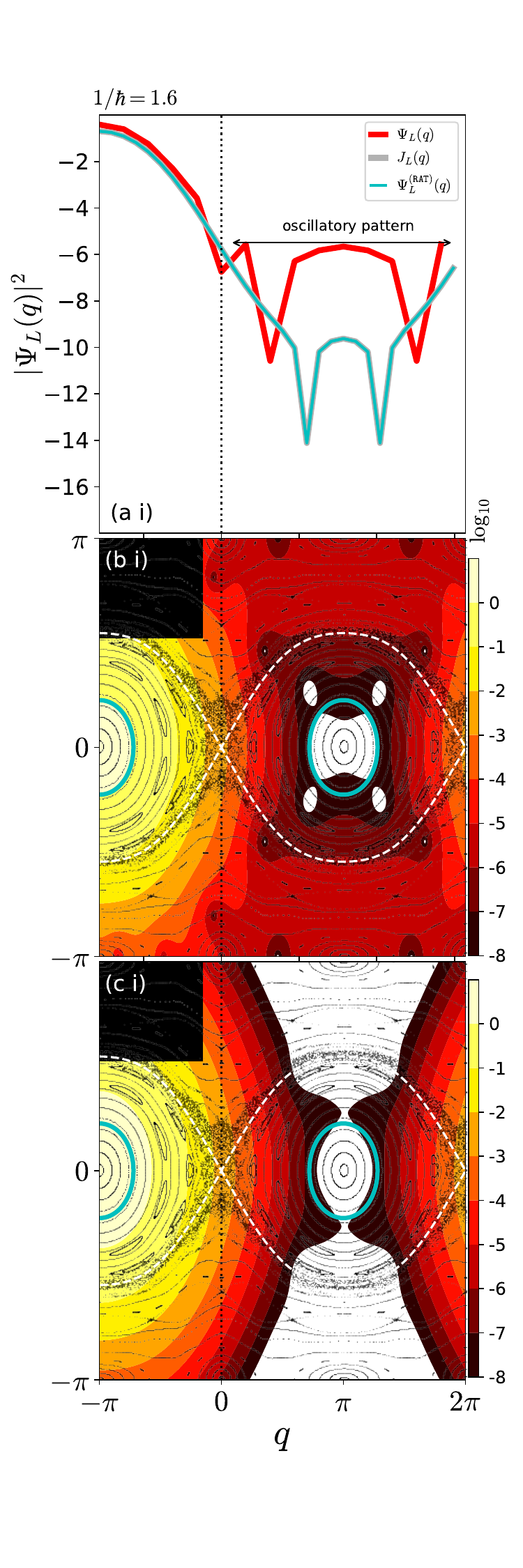}
  \includegraphics[width=0.245\textwidth,trim={1.5cm 2cm 0 0},clip]{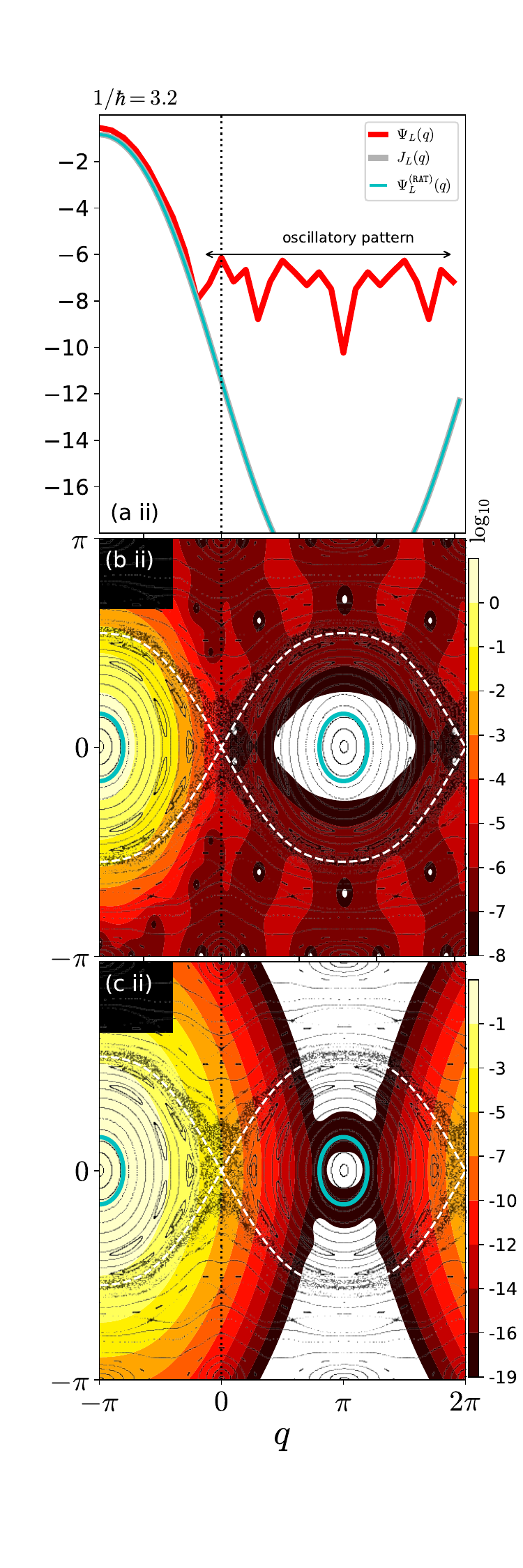}
  \includegraphics[width=0.245\textwidth,trim={1.5cm 2cm 0 0},clip]{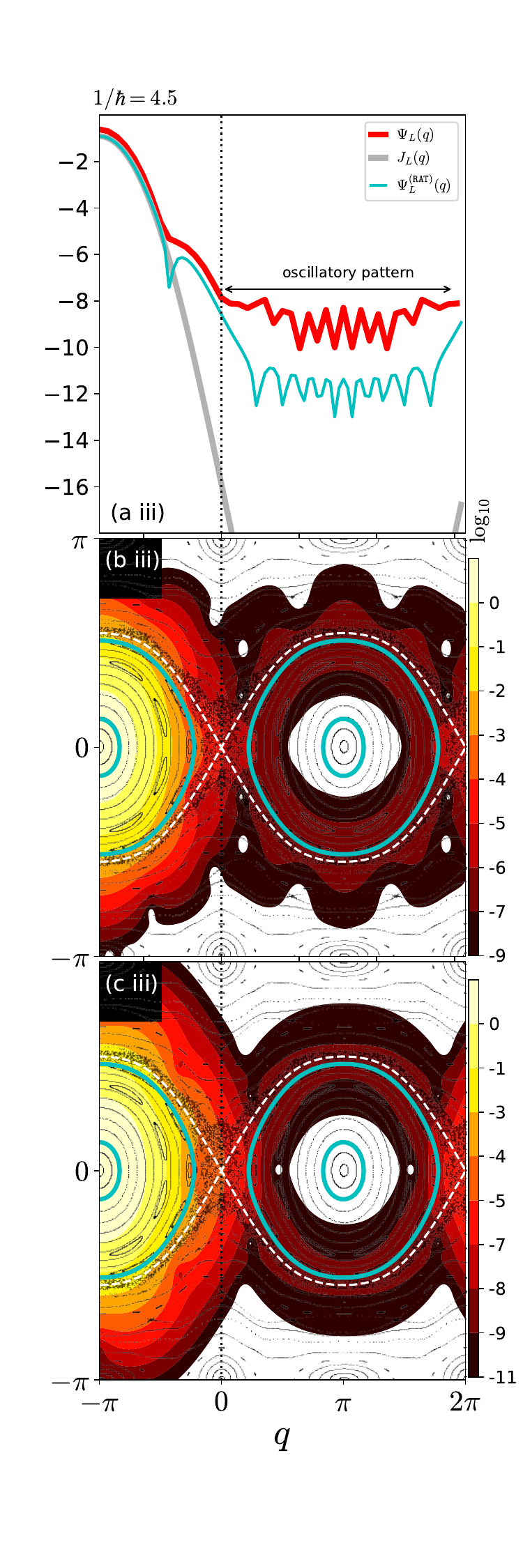}
  \includegraphics[width=0.245\textwidth,trim={1.5cm 2cm 0 0},clip]{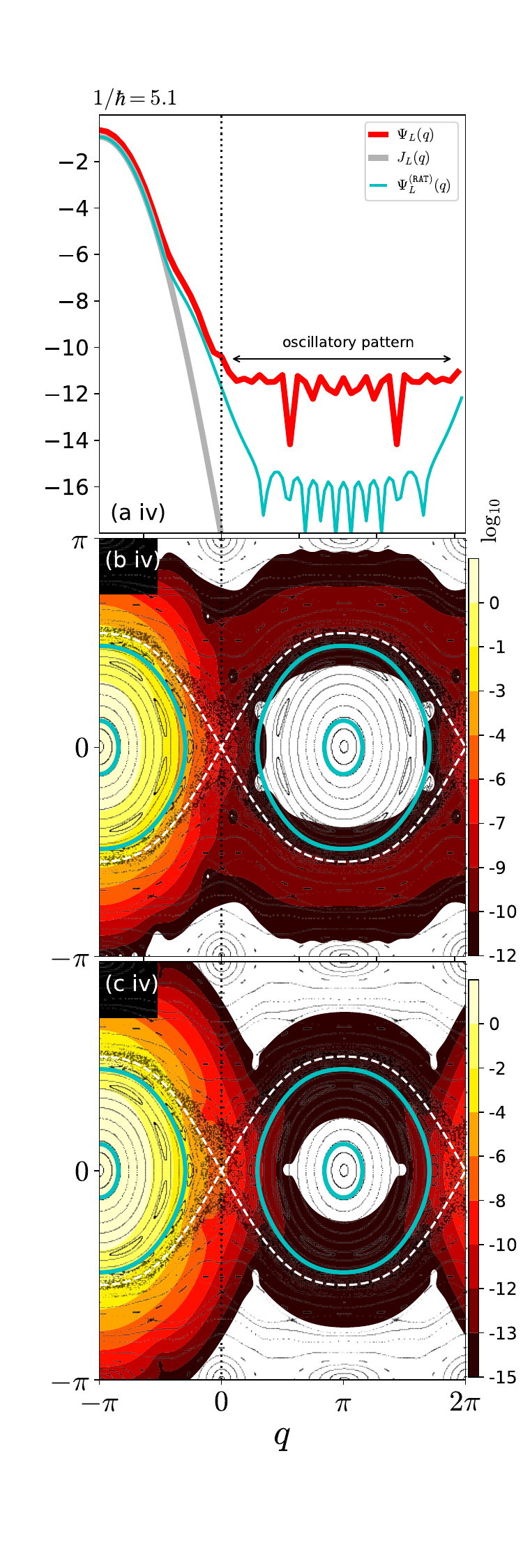}
  }
  \caption{(color online)\label{fig:L-state-hbar}
    (a) The red, cyan, and gray solid curves represent the exact wave function $\Psi_L(q)$,
    the RAT wave function $\Psi_L^\mathtt{(RAT)}(q)$, and
    the integrable one $J_L(q)$ in the $q$ representation, respectively,
    for (i) $1/\hbar=1.6$, (ii) $1/\hbar=3.2$, (iii) $1/\hbar=4.5$, and (iv) $1/\hbar = 5.1$.
    The plots of the Husimi representation for (b) $\Psi_L(q)$ and (c) $\Psi_L^\mathtt{(RAT)}(q)$ are in logarithmic scale.
    The energy contour $E_{n+kr}$ ($k=0, 1$) and $E=\eps$ are indicated by the cyan solid curves.
    The white dashed curve is included as a guide.
    The effective Planck cell is indicated by the black box in the top left corner of each panel.
  }
\end{figure*}

To obtain the tunneling splitting, we can use the following formula based on the Herring-Wilkinson formula
\begin{equation}\label{eq:rat_spl}
\Delta \E_n^\mathtt{(RAT)} = \Delta E_n + \sum_{0\le k\le k_c} |B_{n,k\ell}|^2 \Delta E_{n+rk}.
\end{equation}
Here $\Delta E_n$ denotes the tunneling splitting of the unperturbed Hamiltonian.

We now perform the RAT calculation and see how it works in a situation where a visible nonlinear resonance appears (see Fig.~\ref{fig:res8_vs_hbar}).
The parameters of RAT are determined through the relations~\cite{eltschka2005,lock2010,schlagheck2011}
\begin{align}
I_{r:s} &= \frac{1}{4\pi} (S_{r:s}^+ +S^-_{r:s}),\\
\sqrt{2m_{r:s} V_{r:s}} &= \frac{1}{16} (S_{r:s}^+ - S_{r:s}^-),\\
\sqrt{\frac{2V_{r:s}}{m_{r:s}}} &= \frac{1}{r^2} \arccos(\tr M_{r:s}/2),
\end{align}
where $S_{r:s}^{+(-)}$ are the phase space region bounded by the outer and inner separatrices for the $r:s$ classical resonance, respectively, and 
$M_{r:s}$ is the monodromy matrix of the stable periodic point associated with the $r:s$ classical resonance.
In the RAT scheme, the unperturbed Hamiltonian can be taken independently of the determination of coupling terms.
Here the truncated BCH Hamiltonian $\hat{H}_\mathrm{eff}$ is used for the unperturbed Hamiltonian.

The recipe of the RAT scheme first requires finding visible nonlinear resonances in the region enclosed by separatrix.
In the present case, the $r:s=8:1$ resonance chain is the lowest resonant condition and most visible, as can be easily seen from Fig.~\ref{fig:res8_vs_hbar}.
Of course, there should be an infinite number of nonlinear resonances buried in the regular region. 
However, if we require that the size of the nonlinear resonances should be comparable to the size of the Planck cell, then the $r:s=8:1$ resonance is the only candidate expected to make a RAT contribution.
According to the RAT prescription, the ground state doublet is coupled as 
$E_0\to E_{16}\to E_{32}\to E_{48}\to \cdots$, if $E_{n+kr} < \epsilon$ holds
\footnote{The quantum numbers $n=16, 32, 48,\ldots$ are equal to $8, 2\times 8, 3\times 8,\ldots$ if they are considered in (reduced) Hilbert space with the same symmetry.
}.
In the case $1/\hbar=4.5$, $8.5$, $12.5,\ldots$, the levels $E_{16}$, $E_{32}$, $E_{48},\ldots$ satisfy the quantum resonance condition (\ref{eq:qres}) with $E_0$.

First, let us focus on the first steeply decaying (instanton) region, labeled (a) in Fig.~\ref{fig:spl_vs_hbar_RAT}.
One can find that the results in Eq. (\ref{eq:rat_spl}), obtained according to the RAT recipe, 
reproduce the splitting curve shown in Fig.~\ref{fig:spl_vs_hbar_RAT}. 
In this case, the upper limit of the sum in Eq. (\ref{eq:rat_spl}) is taken as $k_c = 0$. 
We should note that the reproducibility of the instanton region depends strongly on how the unperturbed Hamiltonian $\hat{H}_\mathrm{eff}$ is constructed, although this issue has not been explicitly addressed in the literature. 
The truncated BCH Hamiltonian $\hat{H}_\mathrm{eff}$ is suitable for this purpose.

Next, we shift our focus to the first plateau region, labeled (b) in Fig. \ref{fig:spl_vs_hbar_RAT}.
In this region, the RAT is obviously unable to reproduce the plateau structure.
This is because the upper limit of Eq.~(\ref{eq:rat_spl}) $k_c$ is still zero in the first plateau region, 
meaning that there is no contribution in the RAT recipe.
Recall that the RAT theory starts with the assumption that the Hamiltonian is 
expressed in terms of action-angle variables.
This restricts the quantum transition within the same action-angle space, 
which means that the RAT scheme can only treat the transition inside the separatrix.
This is a fundamental limitation of the integrable approximation framework. 
One cannot go beyond the separatrix with a single pair of action-angle variables.
One could alternatively say that in the plateau region the state giving the coupling to the ground state via the $r:s=8:1$ 
resonance is located beyond the separatrix.
This can also be justified by the fact that the most dominant mode of contribution decomposition in the plateau region 
is beyond the separatrix (see Ref.~\cite{hanada2015}).

It would be worth examining the characteristic of the wave function since the splitting is well approximated by the amplitude of the associated wave function $\ket{\Psi_{L(R)}}$ at $q=0$.
As shown in Figs.~\ref{fig:L-state-hbar}(a i)-\ref{fig:L-state-hbar}(a ii), 
the result of the RAT wave function $\ket{\Psi_n^\mathtt{(RAT)}}$ is equals to $\ket{J_n}$ in the first steeply decaying and plateau region due to $k_c=0$.
However, as clearly shown in Fig.~\ref{fig:L-state-hbar}(a ii),
there is a significant difference between $\ket{\Psi_L}$ and $\ket{\Psi^\mathtt{(RAT)}_L}$ in the region $0<q<2\pi$. 
The Husimi representation of these wave functions tells us that the tunneling tail of $\ket{\Psi_L}$ has a large amplitude in the transversal KAM curve region, which means the tunneling occurs across the separatrix.
This is evidence implying that the coupling beyond the separatrix is
responsible for the formation of the plateau in the $\Delta \E_0$ vs $1/\hbar$ plot.

The value of $k_c$ switched to 1 when the second steeply decaying region starts. 
In other words, the perturbation (second) term in Eq.~(\ref{eq:rat_spl}) starts to contribute to $\Delta \E_n^\mathtt{(RAT)}$.
As demonstrated in Fig.~\ref{fig:spl_vs_hbar_RAT},
the result of the RAT calculation shows rather good agreement with the exact one in the second steeply
decaying region.
This implies that the amplitude of the wave function at $q=0$ is also well predicted
by the RAT calculation, which can be indeed verified in Figs.~\ref{fig:L-state-hbar}(a iii) and \ref{fig:L-state-hbar}(a iv).
We notice that the RAT calculation reasonably reproduces the wave function
for the region $q \le 0$.
The RAT prediction typically shows artificial jumps (see Fig.~\ref{fig:spl_vs_hbar_RAT}). 
This is a known problem, as pointed out in Ref.~\cite{eltschka2005}, that is due to the discontinuous incrementation of $k_c$ 
with varying $\hbar$.

At first glance, the RAT scenario captures the tunneling mechanism 
at least for the second steeply decaying region, that is, the tunneling coupling is enhanced via the classical nonlinear resonance. 
However, as will be argued below, the mechanism described 
by the RAT scenario only works in a limited situation.
We will closely examine the assumptions implicit in the RAT theory and 
carefully consider the mechanism underlying the tunneling process.

Before going into more detail, we would like to emphasize 
that the limitation of the RAT prediction is already manifested in the wave function profile for the $q > 0$ region. 
As shown in Figs.~\ref{fig:L-state-hbar}(a iii) and \ref{fig:L-state-hbar}(a iv), 
the RAT wave function does not predict the exact one in $q > 0$ even in the second steeply decaying region, 
and agreement is achieved only for the region $q \le 0$. 
The RAT theory fails to predict the tunneling transition across the separatrix. 
As can be seen in Fig.~\ref{fig:L-state-hbar}(a iv), 
the region of the wavefunction showing oscillatory patterns has a support
outside of the separatrix [see Fig.~\ref{fig:L-state-hbar}(b iv)]. 
Since the RAT scheme only considers the transition within 
the separatrix, it can never predict the oscillatory region.
According to the Herring-Wilkinson splitting formula, we should recall that 
$\Delta \E_n$ reflects the value of the wave function only at $q=0$. 
Therefore, if $q=0$ is included in the region showing oscillatory patterns, 
the RAT calculation fails to predict $\Delta \E_n$, 
and such regions appear periodically in the $\Delta \E_n$ vs $1/\hbar$ plot.

\subsection{\uppercase{Origin of persistent coupling}}\label{sec:persistent}

\begin{figure}
	\centering
	\includegraphics[width=0.5\textwidth]{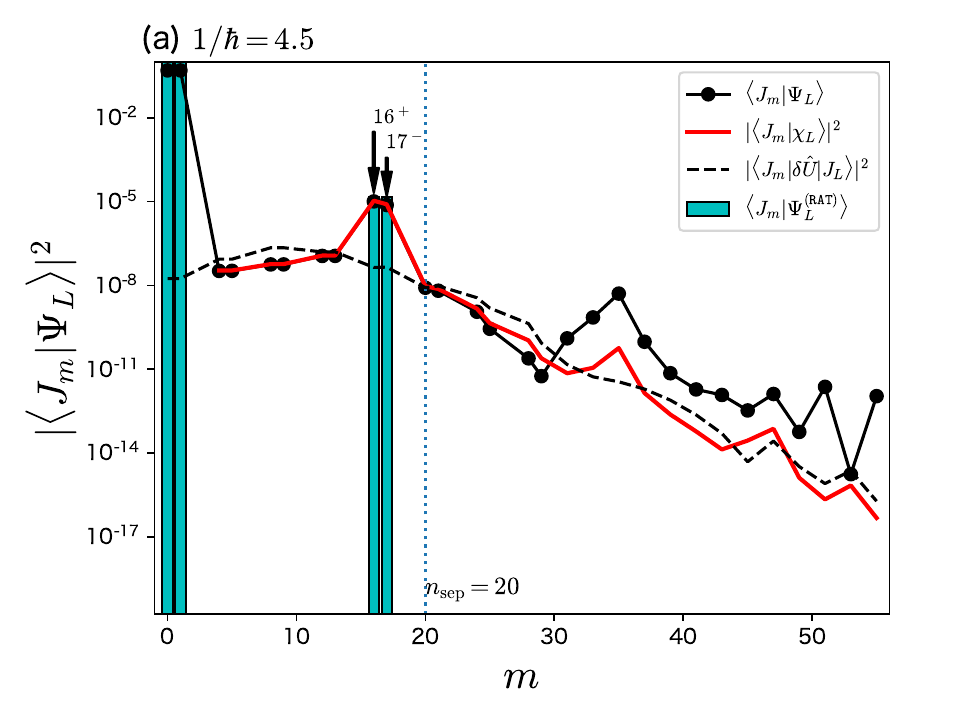}
	\includegraphics[width=0.5\textwidth]{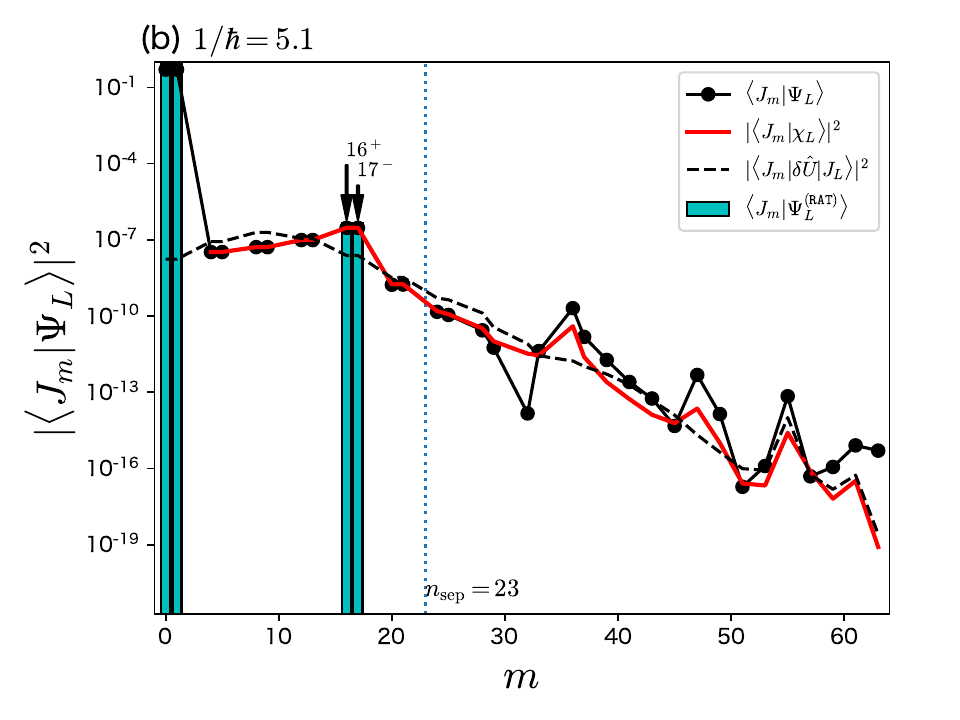}
	\caption{(color online)\label{fig:matele_vs_hbar}
			The solid black and red lines indicate the expansion coefficients for the exact wave function $\bracket{J_m}{\Psi_L}$ and
			the wave function $\bracket{J_m}{\chi_L}$ for (a) $1/\hbar=4.5$ and (b) $1/\hbar=5.1$.
			The black dashed curve represents $\bra{J_m}\delta \hat{U}\ket{J_L}$.
			The cyan bar represents the expansion coefficient $\bracket{J_m}{\Psi_L^{(\mathtt{RAT})}}$.
			The black arrow indicates the resonant (third) state (see also Fig.~\ref{fig:spl_vs_hbar_RAT}).
		}
\end{figure}

Here we focus on the spike observed in the second steeply decaying region labeled (c) in Fig.~\ref{fig:spl_vs_hbar_RAT}.
At $1/\hbar=4.5$, as shown in Fig.~\ref{fig:spl_vs_hbar_RAT}, the spike appears in the splitting curve as a result of the interaction between the ground-state doublet and the third states labeled by $16^+$ and $17^-$ (see Fig.~\ref{fig:spl_vs_hbar}). 
The quantum resonance can be associated with the $r:s=8:1$ classical resonance since the third level is located on the opposite side of the ground-state doublet across the classical resonance [see Fig. \ref{fig:res8_vs_hbar}(a)].

This is exactly the expected situation in the RAT scheme, where the underlying classical resonance is directly linked 
to a quantum resonance and responsible for generating spikes in the splitting curve.
The RAT theory provides a local Hamiltonian for the situation observed here 
by means of the classical canonical perturbation analysis.
Note that this correspondence was analyzed in~\cite{ozorio1984} and later argued in the context of the RAT theory~\cite{brodier2002}.
In the latter paper, the instanton description is applied by constructing the local pendulum Hamiltonian. 
References \cite{wisniacki2011,wisniacki2015} have actually demonstrated that such correspondence 
holds in relation to the RAT theory.

Now we move our attention to the region away from the spike. 
Here it is important to recall that, even in the region without spikes, 
we have kept the couplings with the third states $16^+$ and $17^-$ 
in the RAT calculation shown above. 
Otherwise, the splitting curve drops down to the instanton level 
(the yellow curve in Fig.~\ref{fig:spl_vs_hbar_RAT}).
However, the RAT theory does not tell us why we should or 
are allowed to keep the coupling with the third states even away from the spikes.
In other words, the RAT contribution has to be introduced persistently, 
even far from an avoided crossing, to reproduce the exact curve. 
Note that there is no {\it a priori} principle within the RAT theory about when to apply quantum perturbations. 
One can switch on and switch off the interaction arbitrarily by hand.


As explained in some literature, the association between classical nonlinear resonances 
and the corresponding quantum states is done in the following way. 
Given a system with two degrees of freedom, whose Hamiltonian 
is a function of two action variables $I_1$ and $I_2$ with a parameter $\lambda$, for simplicity, 
if quantum levels degenerate at $\lambda = \lambda_0$, 
the situation can  be expressed semiclassically as $H(I_1,I_2,\lambda_0) = H(I_1',I_2',\lambda_0)$. 
Provided $|I_n - I'_n| \ll1 ~(n=1,2)$, we can expand $H(I_1',I_2',\lambda_0)$ 
around $I'_n = I_n$ to obtain
\begin{align}
\nonumber
 H(I_1',I_2',\lambda_0) &= H(I_1,I_2,\lambda_0)  \\
 \label{eq:expansion}
 &+ (I_1-I'_1) \omega_1 + (I_2-I'_2) \omega_2 + \cdots, 
\end{align}
where $\omega_n = \partial H/\partial I_n~(n=1,2)$. 
Now assuming the semiclassical quantization condition $I_n =( m + \alpha_n/4) \hbar$,
where $\alpha_n$ is the Maslov index, 
we find the condition at $\lambda = \lambda_0$,
\begin{align}
r \omega_1= s\omega_2, 
\end{align}
where $r =m_1 - m'_1$ and $s = m_2 - m'_2$, 
which is exactly the classical resonance condition. 
Therefore, the degeneracy or interaction of two states implies the existence of a resonance in the corresponding classical system.
The interaction generally makes the degenerated levels lift slightly to give an avoided crossing. 
In other words, the presence of avoided crossings may be linked to the classical nonlinear resonances. 

It should be recalled, however, that 
not all avoided crossings are necessarily associated with
classical nonlinear resonances, since avoided crossings also occur in one-dimensional systems~\cite{deunff2010,schlagheck2011}, where resonances do not exist.
Therefore, the correspondence between avoided crossings and classical nonlinear 
resonances is a question to be investigated and is still an ongoing topic~\cite{ramachandran1993influence,wisniacki2011,wisniacki2015,arranz2021correspondence,keshavamurthy2005a}. 

The basic idea of the RAT theory is that a nonlinear resonance mediates the two states straddling the classical resonance. 
Therefore, the invariant manifolds supporting the two states should be located on opposite sides of the resonance 
under consideration. Otherwise, the classical resonance cannot produce the tunneling coupling.
When varying $\epsilon$ or $1/\hbar$, 
one can realize such a situation only at a certain value of $\epsilon$ or $1/\hbar$ 
[see Fig.~\ref{fig:res8_vs_hbar}(a) and Fig.~\ref{fig:spl_vs_hbar_RAT}]. 
If the doublet associated with the avoided crossing interacts with a third state,
the spike is created precisely at this moment. 
However, RAT persistently introduces the coupling associated with the $1:8$ classical resonance even if there exist no $1:8$ classical resonance in between the levels $E_0$ and $E_{16}$ 
if the resonant condition is not satisfied [cf. Figs. \ref{fig:res8_vs_hbar}(b) and \ref{fig:spl_vs_hbar_RAT}].
There is no justification for keeping the coupling in the off-quantum-resonance situations.

It is also important to note that the same mechanism involving an avoided crossing and 
a third state works not only in the nonintegrable system but also in the completely integrable system.
The spikes in the tunneling splitting plot appear there, 
as pointed out in Ref.~\cite{schlagheck2011} and actually demonstrated in Ref.~\cite{deunff2010}. 
Thus, one cannot say that the enhancement invoked by instantaneous spikes 
implies the quantum manifestation of classical nonintegrability.

These arguments immediately raise the question of why the RAT calculation shown in Fig.~\ref{fig:spl_vs_hbar_RAT} 
reasonably predicts the tunneling splitting in the second steeply decaying region.
A key to understanding this is the existence of the broadly spread interactions 
observed in Fig.~\ref{fig:matele_vs_hbar}.
As mentioned above, the broad peak is robust against
the change of $1/\hbar$ and $\epsilon$. 
It survives persistently so that the interaction is maintained even away from the peaks or avoided points.
As a result of such a coupling signature,
the coupling calculated based on the RAT theory 
could be replaced by another component in the broad peak, leaving the result unchanged.
We have already confirmed that this is indeed possible by performing the absorption experiment.
Even if we remove the coupling associated with the RAT calculation, the decay region 
does not drop to the instanton level and remains the same. 
This clearly shows that the RAT coupling is not a necessary condition 
for reproducing the exact calculation.

We can provide additional evidence showing that 
the coupling strength can be reproduced by the RAT 
and that it is not specific to the coupling obtained by the RAT.
As displayed in Fig.~\ref{fig:matele_vs_hbar}(a),
the modes $16^+$ and $17^-$, which are dominant when spikes are generated,
are no longer dominant modes; only one of them is when the parameter is away from the spike 
[see Fig. \ref{fig:matele_vs_hbar}(b)].

The enhancement due to the presence of spikes or avoided crossings
and the persistent enhancement supported by
widespread couplings should be distinguished as different mechanisms.
From the perspective of perturbation theory,
one can say that the former occurs as a result of the
near degeneracy of the energy denominator, while the latter
comes from the nature of transition matrix elements~\cite{hanada2019}.

A crucial issue is therefore reduced to explaining why the persistent enhancement found in the
$\Delta \E_n$ vs $\eps$ and $\Delta \E_n$ vs $1/\hbar$ plots emerges.
In Ref.~\cite{hanada2019}
we investigated in detail and found a singular nature of the broadly extended couplings,
which also supports that the coupling signature is quite different from that predicted by RAT theory.
A deeper understanding of this nontrivial coupling signature
should be explored based on the semiclassical analysis~\cite{shudo2009a,shudo2009b}.
However, this is beyond the scope of the present paper~\cite{koda2023}.

\section{Conclusion and Discussion}
\label{sec:Conclusion}

As reported in previous studies, the width of the tunneling splittings 
deviates from that predicted by the integrable approximation~\cite{roncaglia1994,bonci1998,brodier2001,brodier2002,mouchet2007,lock2010,keshavamurthy2011,hanada2015,hanada2019}. 
Once it deviates from the integrable prediction, it persists even if  
the inverse Planck constant~\cite{hanada2015} is varied or, as shown in Sec.~\ref{sec:spl_vs_eps}, 
some system parameter is changed. 
It would therefore be reasonable to regard the observed phenomena as a manifestation of the nonintegrability of the system. 
The origin of such enhanced tunneling should go back to the underlying classical mechanics. 
The task requires the semiclassical formulation of specifying 
individual energy levels for mixed-type nonintegrable systems, but unfortunately it is unavailable. 

In this paper, using the integrable basis constructed from the BCH expansion, we 
first studied the mechanism of strong and persistent enhancement, 
by observing the response to absorbing perturbation, and the nature of wave functions.
We found that the deviation of tunneling splittings from the integrable prediction 
is observed when we plot the splitting not only as a function of the inverse 
Planck constant $1/\hbar$ but also as a function of the perturbation strength 
$\eps$ of the system. 
The latter plot allows us to see the origin of the enhancement 
more directly, as discussed in Sec.~\ref{sec:spl_vs_eps}. 

If one sweeps the system parameter, the splitting width is enlarged and shows a spiky peak. 
Spikes appear as a result of the collision of a doublet with a third state to produce 
an avoided crossing in the energy space.
The relationship between these spikes and the enhancement has been a matter of discussion since the notion of 
chaos-assisted tunneling was proposed~\cite{bohigas1993,tomsovic1994}. 
It should be recalled that spikes can also be seen even in integrable systems, 
so the existence of spikes is not a property shared only by nonintegrable systems. 

Here we have claimed that the origin of spikes and the persistent enhancement of tunneling splittings should be distinguished. 
This was shown by applying the absorbing perturbation technique~\cite{hanada2015}: Even after removing interactions associated with avoided crossings, the enhancement of tunneling couplings remains. 
This strongly suggests that the type of interactions supporting 
the persistent enhancement is of a long-range nature in the energy space. 
In addition to the energy states forming an avoided crossing in question, 
also many other levels are involved in the persistent enhancement.

With the help of wave-function-based analysis, we revealed that the coupling across the separatrix is responsible for the persistent enhancement of tunneling couplings. The Herring-Wilkinson formula allowed us to analyze 
the behavior of tunneling splittings as a function of the perturbation parameter $\eps$. 
With this tool, it was shown that tunneling splittings are well reproduced by the value of wave function at the central unstable fixed point. At the same time, it was pointed out that the wavefunction in other regions has richer information than the splitting width. 

Local modes obtained by the superposition of the states forming a doublet provide information 
about which components contribute to the splitting enhancement. 
It is essential that the coupling with the outer rotational domains is always present 
even in the parameter regime where the integrable approximation works. 
Moreover, nothing seems to happen in the splitting plot. 
The strong enhancement of tunneling splittings occurs when the contribution 
from the outer region exceeds that predicted by the integrable approximation. 
After the transition, the splitting width increases as a function of the perturbation parameter 
and then decreases again. It has been shown that the coupling with the outer regions is always dominant. 
As a result, the wavefunction exhibits a plateau with oscillatory patterns, whose 
signature contrasts sharply with the local modes obtained by the integrable approximation. 

Based on the analysis of splittings and wave functions in parameter space, 
we revisited the plot of the splitting against the inverse Planck constant. 
In particular, we examined the validity of the RAT scenario as an explanation for the persistent enhancement of tunneling couplings. 
Since the original idea of the RAT theory was to evaluate the tunneling coupling based on the integrable approximation,  i.e., the coupling between symmetrically invariant manifolds via a nonlinear resonance,
the coupling cannot be introduced between the regions that are not connected by the integrable approximation. 
The transition across the separatrix, or equivalently the coupling with 
outer states, cannot be captured by the type of coupling assumed in the RAT theory. 
As demonstrated in Sec.~\ref{sec:spl_vs_hbar}, the situation following the original spirit of 
the RAT theory does exist and the proposed prediction has been shown to work. 
However, such a situation is limited, occurring only around spikes of splitting. 

On the other hand, as explained above, the strong enhancement 
observed for the first time just after the transition from the instanton 
to the plateau region appears due to the coupling with outer states. 
This situation is obviously beyond the scope of the RAT theory. 
To reproduce the exact quantum behavior, one has to maintain the coupling 
``by hand"; otherwise the splitting curve will drop to the curve predicted 
by the integrable approximation. The RAT couplings nevertheless provide 
a reasonable prediction because the coupling for the exact state is broadly spread, not localized 
around a particular resonance state. The existence of broad coupling 
thus explains why keeping the RAT coupling yields a reasonable result, 
even though it is inconsistent with the original idea: incorporating only 
the coupling associated with specific nonlinear resonances. 

In addition, in the periodically perturbed system, 
the spikes of the splitting curve $\Delta \E_n$ appear due to 
the quantum resonance condition (\ref{eq:qres}). 
This results in a one-to-one correspondence between the spikes and the avoided crossings. 
Each spike can be interpreted as a single- or multiphoton (quantum) absorption process except 
for the case $k=0$ in the quantum resonance condition (\ref{eq:qres}).
It could be possible to associate the quantum resonances for the case $k=0$ with 
classical resonances, as discussed in time-independent two-dimensional Hamiltonian systems \cite{ozorio1984,heller1995}. 
On the other hand, in \cite{lock2010,schlagheck2011} the authors tried to link the quantum resonances for $k \neq 0$ to some classical resonances. 
However, in order to develop an argument analogous to that made in \cite{ozorio1984,heller1995} for periodically perturbed systems, one should first establish the relation between avoided crossings and classical resonances in the case of $k=0$.

Furthermore, as explained in Sec.~\ref{sec:persistent}, even in the parameter regime 
where the steep decay appears in the splitting vs the inverse Planck constant plot, the widespread coupling supports an overall bottom-up from the integrable curve because the corresponding regions are already beyond the first transition point from the instanton to the plateau regime in the splitting vs the perturbation parameter $\eps$ plot. 

In this paper the nature of the tunneling couplings has been investigated 
by observing the matrix elements under the BCH basis or wave-function-based arguments. 
It turns out that the origin of the persistent enhancement can be reduced 
to the existence of the interaction over many states and the coupling across the separatrix.
Recall that the role of the separatrix in the system with a simple parabola potential was studied in the Wigner representation \cite{balazs1990}. 
The type of tunneling studied there is different from the tunneling across the separatrix discussed in this paper, but the nonlocal nature of the wave functions induced by the presence of the separatrix could be shared in our situations. 
In any case, 
as already emphasized, the link to the underlying classical mechanics is 
still missing. The present observation has to be reformulated in the language of complex classical dynamics.

One aspect consistent with our conclusion here is that complex orbits can wander anywhere in the complex space in an ergodic way, even though the real phase-space is divided into regular and chaotic components. 
This fact has so far been rigorously proved only for polynomial maps~\cite{bedford1991,bedford1991b,bedford1992a,bedford1992b}, 
but numerical results suggest that it holds more generally~\cite{lazutkin1997homoclinic,shudo2011complex,koda2022complexified}. 
Ergodicity in the complex plane, more precisely in the Julia set, means 
that all orbits initially located in the regular region move not only 
inside the separatrix but also into the outer region. 
In other words, no matter how close to 
the real plane the initial conditions are, complex orbits can go over the separatrix. 
It is very likely that the tunneling couplings can appear via the complex space. 
For this reason, it would be a possible scenario that the tunneling coupling across the 
separatrix leading to the strong enhancement arises from the ergodic nature of the 
dynamics in the complex plane~\cite{koda2023}. 
If this is true, we should say that chaos in complex space produces 
an anomalous tunneling transport that is completely absent in integrable systems.

\section{Acknowledgment}

The authors are grateful to Normann Mertig for his stimulating discussions on resonance-assisted tunneling and for offering explicit parameters used in the paper. 
Y.H. thanks Kin'ya Takahashi for encouraging this work. 
This work has been supported by JSPS KAKENHI Grant No.~17K05583 and 22H01146. Y.H. acknowledges the support of a Showa University Research Grant for Young Researchers.

\appendix

\section{The splitting formula for a Floquet system}\label{app:spl}

In this appendix, we derive the splitting formula for the Floquet system. 
The derivation essentially follows the idea used in Ref. \cite{creagh1998}.

Here we consider the time-dependent Schr\"odinger equation, 
\begin{equation}\label{app2:schr}
 \hat{H}(p,q,t) = - \frac{\hbar^2}{2}\pdo{^{2}}{q^{2}} + V(q,t),
\end{equation}
with a time-periodic potential $V(t+\tau)=V(t)$, where $\tau$ is the period.
Let $\omega=2\pi/\tau$ be the frequency of the periodic perturbation. 
Let $\mathcal{H}(\Omega)$ be the associated Hilbert space. 
Here $\Omega$ denotes the domain on which the wave function acts.

The Floquet theorem guarantees the existence of a solution satisfying 
\begin{align}
 \psi_{(n,m)}(q,t) &= e^{-i\E_{(n,m)} t/\hbar} u_{(m,n)} (q,t),\\
 u_{(n,m)}(q,t+\tau) &= u_{(n,m)}(q, t),\\
 \hat{K}u_{(n,m)}(q,t) &= \E_{(n,m)} u_{(n,m)}(q,t),
\end{align}
where $\hat{K}=\hat{H}(t) -i\hbar\partial_t$ is a Hermitian operator \cite{sambe_steady_1973,holthaus_floquet_2015} and 
\begin{align}
 \E_{(n,m)} &= \E_n + m\hbar\omega,\\
  u_{(n,m)}(q,t) &= u_n(q,t)e^{im\omega t}.
\end{align}
where $m\in\mathbb{Z}$.
The 2$\pi$ modulus of the time-evolution operator gives rise to multivalued quasienergies, 
which is known as the Brillouin structure.
The operator $\hat{K}$ acts on the extented Hilbert space denoted by $L^2([0,\tau])\otimes\mathcal{H}(\Omega)$.
Let $f$ and $g$ be the states in the extended Hilbert space and the associated inner product be 
\begin{equation}
\< f| g\>  = \frac{1}{\tau}\int_{0}^{\tau}\int_\Omega f^\ast(q,t) g(q,t) dx.
\end{equation}
In the following discussion, we limit ourselves to the first quasi-energy Brillouin zone $m=0$, and omit the suffix $m$ 
hereafter. 
We further assume the Hamiltonian has a symmetry as $H(q,t)=H(-q,t)$, 
which could lead to a set of congruent torus in the corresponding classical phase space.

We consider here the ground-state doublets $u_0(q,t)$ and $u_1(q,t)$ in the same sense in the main text, 
and define the associated tunneling splitting as $\Delta \E = |\E_{1} - \E_{0}|$. 

The localized states on left (right) tori are introduced as 
\begin{equation}
\ket{L(R)} = (\ket{u_0} \pm \ket{u_1})/\sqrt{2},
\end{equation}
whose the quasi-energies are degenerated to give $\tilde{\E} = (\E_1 + \E_0)/2$.
By using these basis, the Floquet Hamiltonian can be expressed as 
\begin{align}
 \hat{K}
 \begin{pmatrix}
  \ket{L}\\
  \ket{R}
 \end{pmatrix}
 =
 \begin{pmatrix}
 \tilde{\E} & \Delta \E/2\\
 \Delta \E/2 & \tilde{\E}
 \end{pmatrix}
 \begin{pmatrix}
  \ket{L}\\
  \ket{R}
 \end{pmatrix}.
\end{align}

Let us introduce time independent Hermitian projection operator \cite{creagh1998} such that
\begin{align}
\hat{\Theta} \ket{L} \approx 0, \qquad\hat{\Theta}\ket{R} \approx \ket{R}.
\end{align}
Tunneling splitting can then be expressed as 
\begin{align}
\Delta \E/2 &\approx \bra{R}[\hat{\Theta}, \hat{K}]\ket{L},\notag \\
& = \bra{R}[\hat{\Theta}, \hat{H}]\ket{L} - \bra{R}[\hat{\Theta}, \partial_{t}] \ket{L} \notag \\
& =\frac{\hbar^{2}}{2\tau} \int_{0}^{\tau} dt \int_{\Omega} dq \Bra{u^{\ast}_{R}\delta u'_{L} - u'^{\ast}_{R}\delta u_{L}}.
\end{align}
where $\delta$ stands for a Dirac delta function.
The second term is equal to zero when we assume $\hat{\Theta}$ is a time-independent.
The first term is evaluated analogously as the time-independent case to give 
\begin{align}
& \<R |  [\hat{\Theta}, \hat{H}]|L\> = \frac{1}{\tau} \int_{0}^{\tau} dt \int_{\Omega} dq\ u^{\ast}_{R}[\hat{\Theta}, \hat{H}] u_{L} \\
&= -\frac{\hbar^{2}}{2\tau} \int_{0}^{\tau} dt \int_{\Omega} dq \Bra{u^{\ast}_{R}\theta u''_{R} - u''^{\ast}_{L}\theta u_{R}},
\end{align}
where the asterisk and the primes represent complex conjugate and the derivative with respect to $q$, respectively.
Here we have done a partial integration based on the fact that 
the boundary contributions to the integral turn out to be zero: 
\begin{align}
u(q, t) = u'(q, t) = 0, \quad q \in \partial \Omega, 
\end{align}
for both periodic and unbounded domains. 
We then obtain the splitting formula for the Floquet system as 
\begin{align}
\Delta \E \approx \frac{\hbar^2}{\tau}\int_{0}^{\tau} dt \Bra{u^\ast_R(0, t)u'_L(0,t) -u'^\ast_R(0,t) u_L(0,t)}.
\end{align}

\label{app:splitting}

\bibliography{ref}
\end{document}